\documentclass[journal,12pt,onecolumn]{IEEEtran}
\usepackage[utf8]{inputenc}
\usepackage[english]{babel}
\usepackage{setspace}
\usepackage{graphicx}
\usepackage{amsmath}
\usepackage{amsthm}
\usepackage{lscape}
\usepackage{latexsym}
\usepackage{amssymb}
\usepackage{bm}
\usepackage{epstopdf}
\usepackage{bbm}
\usepackage{cite}
\usepackage{url}
\usepackage{color}
\usepackage[caption=false,font=footnotesize]{subfig}
\usepackage{graphicx}
\usepackage{lmodern}
\usepackage[euler]{textgreek}

\newtheorem{remark}{Remark}
\newcommand\SNR{\mathrm{SNR}}

\newcommand\blfootnote[1]{%
	\begingroup
	\renewcommand\thefootnote{}\footnote{#1}%
	\addtocounter{footnote}{-1}%
	\endgroup
}

\usepackage{tikz}
\usepackage{tikzsymbols}
\usepackage{float}
\usetikzlibrary{shapes.geometric, arrows, positioning, fit}

\title{DeepJSCC-$f$: Deep Joint Source-Channel Coding \\of Images with Feedback}

\author{
   \IEEEauthorblockN{David Burth Kurka and Deniz Gündüz}
   \\
    \IEEEauthorblockA{Imperial College London
    \\\{d.kurka, d.gunduz\}@imperial.ac.uk}}
\date{}

\begin{document}

\maketitle

\begin{abstract}
    We consider wireless transmission of images in the presence of channel output feedback. From a Shannon theoretic perspective feedback does not improve the asymptotic end-to-end performance, and separate source coding followed by capacity-achieving channel coding, which ignores the feedback signal, achieves the optimal performance. It is well known that separation is not optimal in the practical finite blocklength regime; however, there are no known practical joint source-channel coding (JSCC) schemes that can exploit the feedback signal and surpass the performance of separation-based schemes. Inspired by the recent success of deep learning methods for JSCC, we investigate how noiseless or noisy channel output feedback can be incorporated into the transmission system to improve the reconstruction quality at the receiver. We introduce an autoencoder-based JSCC scheme, which we call DeepJSCC-$f$, that exploits the channel output feedback, and provides considerable improvements in terms of the end-to-end reconstruction quality for fixed-length transmission, or in terms of the average delay for variable-length transmission. To the best of our knowledge, this is the first practical JSCC scheme that can fully exploit channel output feedback, demonstrating yet another setting in which modern machine learning techniques can enable the design of new and efficient communication methods that surpass the performance of traditional structured coding-based designs.

\end{abstract}

\blfootnote{This work was supported by the European Research Council (ERC) through project BEACON (No. 677854).}

\section{Introduction}
One of the fundamental results in information theory is Shannon's separation theorem \cite{Shannon:1948}, which establishes that when transmitting a source signal over a noisy channel there is no loss in optimality by first compressing the source samples into bits, ignoring the channel statistics, and then transmitting the compressed bits over the
channel using an optimal channel code, designed independently of the source statistics. This modular design principle has had a tremendous impact on research as well as practice, reinforcing the notion of layering in communication networks. The application layer (which processes the information source) does not need to know the details of the physical layer (which deals with modulation and channel coding), and vice versa. However, despite its huge impact, optimality of separation holds only under unlimited delay and complexity; and even under these assumptions, it breaks down in multi-user scenarios \cite{Shannon:1961, Gunduz:IT:09}, or for non-ergodic source or channel distributions \cite{Vembu:IT:95, Gunduz:IT:08}. 

Despite its suboptimality, almost all communication systems today are designed based on the separation approach. This is due not only to the modularity of the separate design, but also to the lack of powerful joint source-channel coding (JSCC) schemes with reasonable coding and decoding complexity. In current systems, we have highly specialized source codes for different type of information sources e.g., JPEG2000/ BPG for images, MPEG-4/ WMA for audio, or H.264 for video, which have been optimized over many decades and many generations of standards, followed by similarly highly optimized channel coding and modulation techniques to be used over different communication channels, e.g., Turbo, LDPC, polar codes. Although there have been many research efforts on joint source-channel coding  techniques, they have mostly focused either on theoretical analysis under some idealistic source and channel distributions \cite{Gastpar:IT:03, Kostina:IT:17, SPA:IT:02, Mittal:IT:02, Hekland:TC:09}, or the joint optimization of the parameters of the components (vector quantizer, index assignment, channel code and modulator) of an inherently separate design  \cite{Cheung:TIP:00, Farvardin:IT:90, Farvardin:IT:91, Skoglund:IT:99a, Kozintsev:TSP:98, Skoglund:IT:99b}. 

Another fundamental result in information theory, again due to Shannon \cite{Shannon:IT:56}, states that feedback does not increase the capacity of a memoryless communication channel. It is not difficult to show that the optimality of separation continues to hold in the presence of feedback; therefore, information theoretically feedback does not help the end-to-end performance of source-channel coding either. On the other hand, feedback is known to improve the error exponents for channel coding \cite{Burnashev:PIT:76, Schalkwijk:IT:66}, and to significantly simplify the design of JSCC schemes, at least in some ideal scenarios \cite{Schalkwijk:IT:67, Kailath:PIEEE:67, Butman:IT:71}. However, there has been limited success in converting these theoretical gains of feedback into practical coding schemes. 

\begin{figure}
	\begin{center}
        \resizebox {0.95\linewidth} {!}%
        {

 \tikzstyle{txt} = [text centered]
 \tikzstyle{box} = [rectangle, rounded corners, minimum width=2.5cm, minimum height=1.5cm, text width=2.5cm, text centered, draw=black]
 \tikzstyle{smallbox} = [rectangle, rounded corners, minimum width=1cm, minimum height=1cm, text width=1cm, text centered, draw=black]
 \tikzstyle{bbox} = [rectangle, thick, minimum width=2.5cm, minimum height=1cm, text width=1.5cm, text centered, draw=black]
 \tikzstyle{arrow} = [thick,->,>=stealth]
 \tikzstyle{fitted} = [draw=gray, thick, dotted, inner sep=0.75em]
 \tikzstyle{sum} = [draw, circle]


\begin{tikzpicture}[node distance=2.3cm]

\node (x) [txt, font=\fontsize{14}{0}\selectfont] {$\bm{x}$};
\node (encoder) [box, right of=x, xshift=1cm, font=\fontsize{12}{12}\selectfont] {Joint Source-Channel Encoder \\ ($f_{\bm{\theta}}$)};
\node (sum) [sum, right of=encoder, xshift=1cm] {\Large$+$};
\node (n) [txt, above of=sum, yshift=-0.7, font=\fontsize{14}{0}\selectfont] {${\bm n}_i$};

\node (decoder) [box, right of=sum, xshift=2cm,  font=\fontsize{12}{12}\selectfont] {Joint Source-Channel Decoder\\($g_{\bm{\phi}}$)};
\node (xhat) [txt, right of=decoder, xshift=1cm, font=\fontsize{14}{0}\selectfont] {$\bm{\hat{x}}$};

\node (delay) [smallbox, below of=sum, yshift=1cm, xshift=1.5cm] {Delay};

\node (sumfb) [sum, below of=sum] {\Large$+$};
\node (nfb) [txt, below of=sumfb, yshift=0.7cm, font=\fontsize{14}{0}\selectfont] {${\bm n}^f_{i-1}$};

\draw [arrow] (x) -- (encoder);
\draw [arrow] (encoder) -- node[above,font=\fontsize{14}{0}\selectfont] {$\bm{y}_i$} (sum);
\draw [arrow] (n) -- (sum);
\draw [arrow] (sum) -- node[above,font=\fontsize{14}{0}\selectfont] {$\bm{z}_i$} (decoder);
\draw [arrow] (decoder) -- (xhat);
\draw [arrow] (nfb) -- (sumfb);

\draw [arrow] (sum) -| (delay);
\draw [arrow] (delay) |- node[below,font=\fontsize{14}{0}\selectfont] {$\bm{z}_{i-1}$} (sumfb);
\draw [arrow] (sumfb) -| node[below,font=\fontsize{14}{0}\selectfont] {$\bm{w}_{i-1}$}  (encoder);

\end{tikzpicture}

}%
	\end{center}
\caption{Communication in the presence of noisy channel output feedback. Source $\bm{x}$ is transmitted over a noisy channel, where after the transmission of each symbol a noisy version of the received signal $\bm{z}_i$ becomes available to the encoder through a feedback link.}
\label{fig:modelfeedback}
\end{figure}
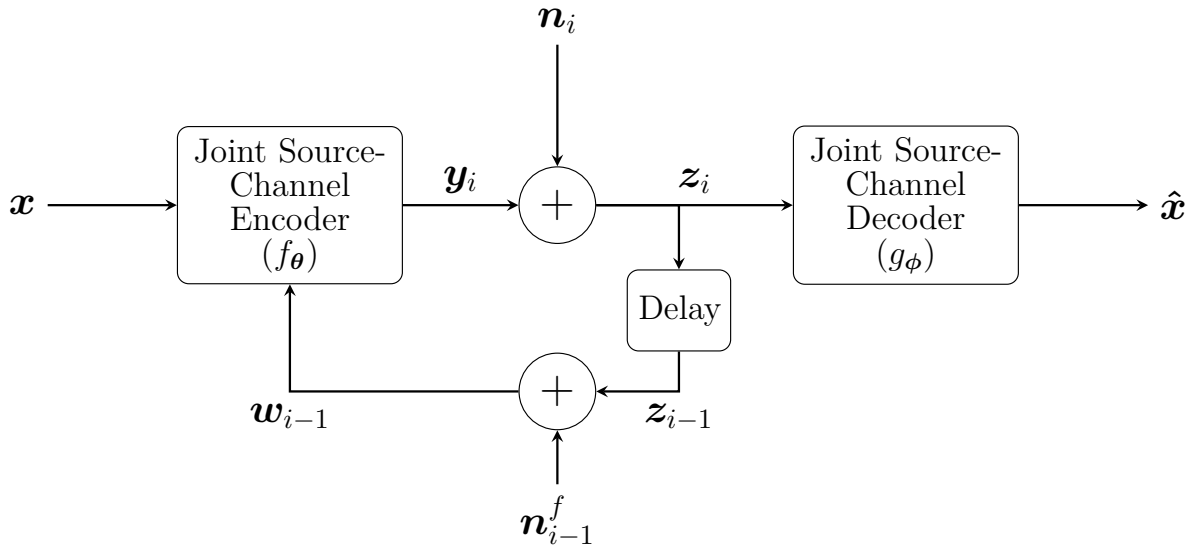

In this paper, our goal is to design practical JSCC schemes that can exploit noisy or noiseless channel output feedback (see Figure \ref{fig:modelfeedback} for an illustration of the system model), building upon the recent success of deep neural networks (DNNs) for various coding and communication problems. The decoder of any known channel code can be treated as a classifier on the noisy channel output. With the growing success of deep learning for various classification tasks, researchers first explored using DNNs to decode existing channel codes, such as linear \cite{Nachmani:JSTSP:18}, turbo \cite{Jiang:DeepTurbo:19}, or polar \cite{Ebada:DeepPolar:19} codes, which provided promising improvements in the error probability. This was later extended to the joint design of the encoder and the decoder using an autoencoder structure \cite{Oshea:TCCN:17}. Despite some  success in designing codes that outperform convolutional codes in the short blocklength regime \cite{Jiang:ICC:19}, extending this success to longer codes has been a challenge due to the exponentially growing codebook size with the blocklength. 

More impressive results have been achieved through DNN-based designs in settings where current codes fall short of the fundamental theoretical limits; for example, for channels that are harder to model, such as optical  \cite{Karanov:JLT:18} and molecular \cite{Farsad:TSP:18} communications, or in settings no practical coding scheme exists even for known channel statistics, such as communication in the presence of channel output feedback \cite{Hyeji:NIPS:2018}. In \cite{Hyeji:NIPS:2018}, authors constructed a code based on recurrent neural networks (RNNs), called \textit{Deepcode}, that can significantly outperform the coding scheme of Schalkwijk and Kailath \cite{Schalkwijk:IT:66} in the case of noiseless feedback, and achieve reasonable reliability even in the case of noisy channel output feedback. Although the authors proposed concatenating Deepcode with other known coding schemes to achieve larger blocklengths, the code itself is still limited to short blocklengths due to training difficulty. Another limitation of Deepcode is that its design is limited to rates $1/r$, for $r=2, 3, \ldots$.

Also related to our work are the parallel advances in data compression using DNNs, particularly for image compression. As opposed to most current image compression standards, such as JPEG, JPEG2000, BPG, that depend on transform coding followed by quantization and entropy coding, researcher have studied autoencoder-based DNNs with impressive results \cite{TodericiICLR2016, TheisICLR2017, Balle:ICLR:17, Johnston_2018_CVPR, BalleICLR2018, Minnen:NIPS:18, Lee:ICLR:19}. These DNN-based compression schemes meet or even surpass standard codecs in various performance metrics, including in terms of subjective perceptual quality \cite{Cheng:perceptual}.

Most related prior work to the current paper are \cite{FarsadICASSP2018, ZarconeDCC2018, Eirina:ICASSP19, Eirina:TCCN:19, Kurka:IZS2020, Kurka:ICASSP19, choi2019neural}, which consider the JSCC problem, and propose autoencoder-based solutions for end-to-end optimization, without explicitly focusing on the compression or the channel coding problems. While \cite{FarsadICASSP2018} focuses on text as the information source and binary channels, and \cite{ZarconeDCC2018} deals with lossy data storage, image transmission over an additive white Gaussian noise (AWGN) wireless channel is studied in \cite{Eirina:TCCN:19, Kurka:ICASSP19, choi2019neural, Kurka:IZS2020}. In \cite{Eirina:TCCN:19} the authors propose a fully convolutional autoencoder architecture, which maps the input images directly to channel symbols, without going through any digital interface, and show that the proposed DeepJSCC architecture not only improves upon the concatenation of state-of-the-art compression and channel coding schemes in a separate architecture \cite{Kurka:IZS2020}, but also provides graceful degradation with channel signal-to-noise ratio (SNR). This latter property, which is common to analog transmission schemes, provides significant benefits compared to digital schemes, which exhibit threshold behaviour; particularly when broadcasting to multiple receivers, or when transmitting over a time-varying channel. DeepJSCC is also shown in \cite{Kurka:ICASSP19} to be almost \textit{successively refinable}; that is, an image can be transmitted in stages, where each stage refines the quality of the previous stages, while this multi-stage structure comes at almost no additional cost.

The current paper builds upon the results of \cite{Eirina:TCCN:19} and \cite{Kurka:ICASSP19}, and proposes an autoencoder architecture for JSCC of images over an AWGN channel with channel output feedback, considering both perfect and noisy feedback scenarios. Note that this is a significantly challenging problem as the neural encoder not only needs to learn how to map the image into the channel input space, while providing robustness against the channel noise, but also needs to learn how to exploit the channel output feedback, which provides side information on the decoder's belief about the source signal. The difficulty of this problem is corroborated by the lack of any practical codes. Only a handful of papers have studied this problem; among the few papers that consider image transmission with feedback, \cite{Chande:98} and \cite{Lu:DCC:99} consider one bit ACK/NACK type feedback, which is used to change the rate of the channel code by sending additional parity bits until an ACK signal is received to indicate correct reception, while \cite{Kafedziski:98} considers channel state information (CSI) feedback for a multicarrier scenario, which is used to decide the source information transmitted per channel together with linear encoders.

\begin{figure}
    \centering
    \includegraphics[width=0.7\linewidth]{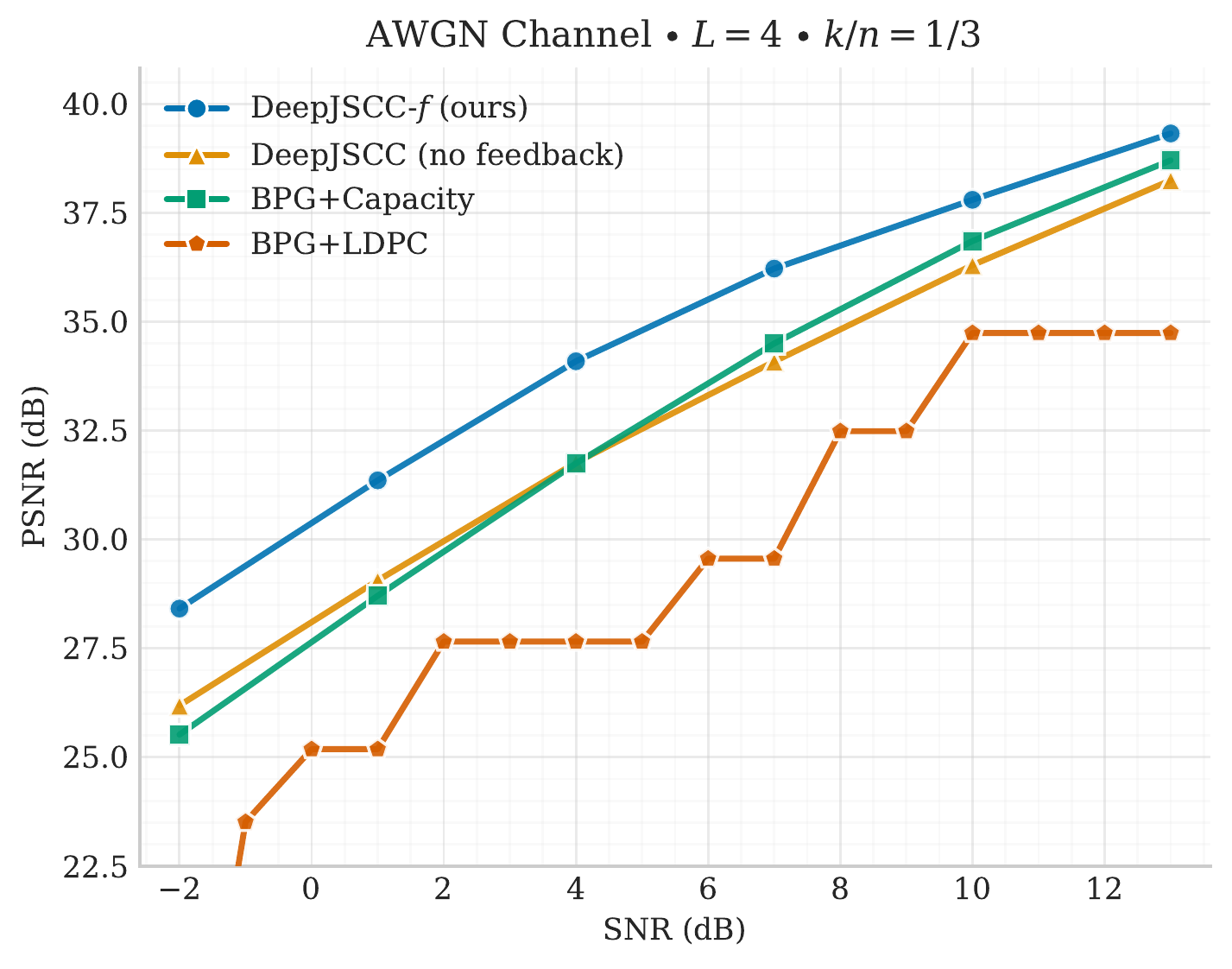}
    \caption{Performance of our algorithm compared to other practical and ideal schemes. Our model using 3 channel feedback transmissions can outperform all other schemes over a wide range of PSNRs.}
    \label{fig:snr4}
\end{figure}

Although the autoencoder-based results for both channel coding and image compression do not yet provide significant improvements over existing standards, we highlight the fact that the latter are results of decades-long intense efforts and expertise of respective engineering communities, while the data-driven DNN techniques have been introduced only a few years ago, and have already achieved impressive results, particularly for the image compression problem. On the other hand, we would like to highlight that for the JSCC of images over wireless channels with feedback, separate design is strictly suboptimal, and we do not have practical JSCC schemes that truly adapt the mapping of the source information to the channel input (and vice versa at the receiver) depending on the channel output feedback. Therefore, the proposed DNN-based coding architecture is the first truly joint design, where the joint mapping of the source to the channel input depends on the feedback signal. Despite the difficulty of the problem at hand, the results presented in this paper are extremely promising, in the sense that, the proposed deep JSCC architecture that exploits the channel output feedback, called \textit{\mbox{DeepJSCC-$f$}}, outperforms any separate source channel coding architecture even if we consider combining state-of-the-art image compression codecs (e.g., JPEG2000 or BPG) with a capacity achieving channel code (despite operating over a finite blocklength). The proposed \mbox{DeepJSCC-$f$} architecture divides the available channel bandwidth into a finite number of $L$ blocks, and transmits the image in $L$ refinement layers, where the transmission over each layer depends on the original image as well as the receiver's reconstruction up to that point. This coding scheme is inspired by the Schalkwijk-Kailath scheme which is designed for a single Gaussian source sample, and is limited to linear encoding of the estimation error of the receiver at each symbol. Instead, we consider the receiver's reconstruction of the original image over blocks, and design a nonlinear encoding and decoding architecture employing convolutional neural networks (CNNs), which improves the quality of the receiver's estimate at each layer. 

We plot in Figure \ref{fig:snr4} the peak SNR (PSNR) for the transmission of images from the CIFAR10 dataset over an AWGN channel with a compression ratio of $1/3$, which refers to the available channel uses per image pixel (please see below for the proper definition). Here we compare the performance of the \mbox{DeepJSCC-$f$} scheme with the DeepJSCC architecture from \cite{Eirina:TCCN:19} and with the concatenation of BPG codec followed by LDPC channel coding. These two schemes do not exploit the feedback link. On the other hand, we also plot the performance obtained by the BPG codec assuming a capacity achieving channel code. Note that, since feedback does not increase the capacity, this corresponds to the best performance that can be obtained by a separation-based scheme that combines BPG compression with any channel coding scheme, including those that exploit the channel output feedback. We observe that BPG followed by a capacity-achieving channel code outperforms DeepJSCC; however, when channel output feedback is utilized, the performance of \mbox{DeepJSCC-$f$} significantly improves.

In summary, we present in this paper the first practical scheme for JSCC of images in the presence of channel output feedback. The proposed architecture is fully convolutional, and thus can be used for input images with different sizes. It employs a layered structure that divides the available channel bandwidth into multiple blocks, and refines the estimate of the receiver with each new layer transmitted over a different channel block. It is shown that there is an optimal number of layers that should be employed. Average reconstruction quality increases gradually with the number of layers up to this optimal value, at the expense of increased complexity and training time. The proposed \mbox{DeepJSCC-$f$} scheme significantly outperforms separation-based schemes even when we assume the combination of state-of-the art image compression codecs followed by a capacity-achieving channel code. The proposed architecture naturally lends itself to variable-length transmission, where the transmitter sends new layers until a certain quality target is met at the receiver. We have shown that the improvement is even more significant in terms of the average required delay with variable-length transmission. Another interesting property of \mbox{DeepJSCC-$f$} is that its performance behaves more like analog transmission than separation-based digital communication, and exhibits graceful degradation with the channel SNR. This is in contrast to digital communication systems, which suffer from the \textit{cliff effect}; that is, the performance breaks down when the channel SNR goes below the target SNR for which the channel code was chosen; and the quality does not increase no matter how high the channel SNR is beyond this target value. Finally, we show that the proposed architecture is robust against variations in the feedback channel quality as well, while most known communication schemes are extremely sensitive to noise in the feedback link as well as mismatches between the targeted and real values of the feedback channel quality.

\section{Problem Formulation}
\label{sec:ProblemFormulation}

We consider the problem of wireless transmission of images using feedback. An input image, represented as a vector of pixel intensities $\bm x \in  \mathbb{R}^n$ is to be transmitted in $k$ uses of a noisy communication channel. In this work, we will consider static as well as fading memoryless complex additive white Gaussian noise (AWGN) channels, modeled as 
\begin{align}
    {\bm z} = h {\bm y} + {\bm n},
\end{align}
where ${\bm y} \in \mathbb{C}^k$ denotes the complex channel vector, ${\bm z} \in \mathbb{C}^k$ the corresponding complex channel output vector, $h \in \mathbb{C}$ is the channel gain that remains constant throughout $k$ channel uses, and ${\bm n} \in \mathbb{C}^k$ is the independent and identically distributed (i.i.d.)\ circularly symmetric complex Gaussian noise vector with zero mean and variance $\sigma^2$. Initially we will consider only AWGN channels, that is, we will set ${\bm z} = {\bm y} + {\bm n}$. When we consider a fading channel, we will assume $h$ to be random with a given distribution, but it will remain constant throughout the transmission of a single image, i.e., quasi-static fading channel model. 
An average power constraint is imposed on the transmitted signal: $\frac{1}{k} \mathbb{E}[{\bm y}^* {\bm y}] \leq 1$, where $^*$ denotes the complex conjugate operation.

We consider the presence of channel output feedback, through which a noisy version of the channel output at the receiver is made available to the transmitter with a unit delay. Hence, the feedback channel is characterized by $\bm{w} = {\bm z} + {\bm n}^f$, where ${\bm n}^f$ denotes the i.i.d. additive complex Gaussian noise vector on the feedback link with zero mean and variance $\sigma^2_f$.

An $(n,k)$ joint source-channel code for this system consists of 
\begin{enumerate}
    \item a sequence of encoding functions $f_i: [0:255]^n \times \mathbb{C}^{i-1} \rightarrow \mathbb{C}$, each of which maps the input image and the channel output feedback until time $i-1$ to channel input at time $i$, that is, ${\bm y}(i) = f_i(\bm x, \bm{w}(1), \dots, \bm{w}(i-1))$, while satisfying the average power constraint, and 
    \item a decoding function $g: \mathbb{C}^{k} \rightarrow \mathbb{R}^{n}$ that maps the received channel output to the reconstructed image $\bm{\hat{x}} = g({\bm z})$.
\end{enumerate}

The image dimension $n$ is also called as the \emph{source bandwidth} and is given by the product of the image's height, width and the number of colour channels. On the other hand, $k$ defines the \emph{channel bandwidth}, and we denote the ratio $k/n$ as \emph{bandwidth ratio}, determining how many channel symbols are available for each source symbol, which may correspond to spectral or temporal resources available. We will concentrate our results on the more practically relevant case of bandwidth compression, i.e., $k < n$, although our techniques are applicable for any bandwidth ratio. 

For given system parameters, the goal is to define a code that optimizes the average system performance, in terms of some specified distortion metric between the input and output images, averaged across possible input signals as well as direct and feedback channel distributions. Note that this model specifies a joint source-channel code that maps a fixed-size input image to a fixed-length channel codeword. On the other hand, thanks to the presence of feedback, we can also consider a variable-length coding setting, where we can set a target distortion value for each image, and minimize the average code length required to meet this target value for each image. This formulation will be considered in Section \ref{ss:variable_rate}.  

\section{\mbox{DeepJSCC-$f$} Architecture}
\label{sec:sysdesign}

\begin{figure}
    \centering
    \includegraphics[width=\linewidth]{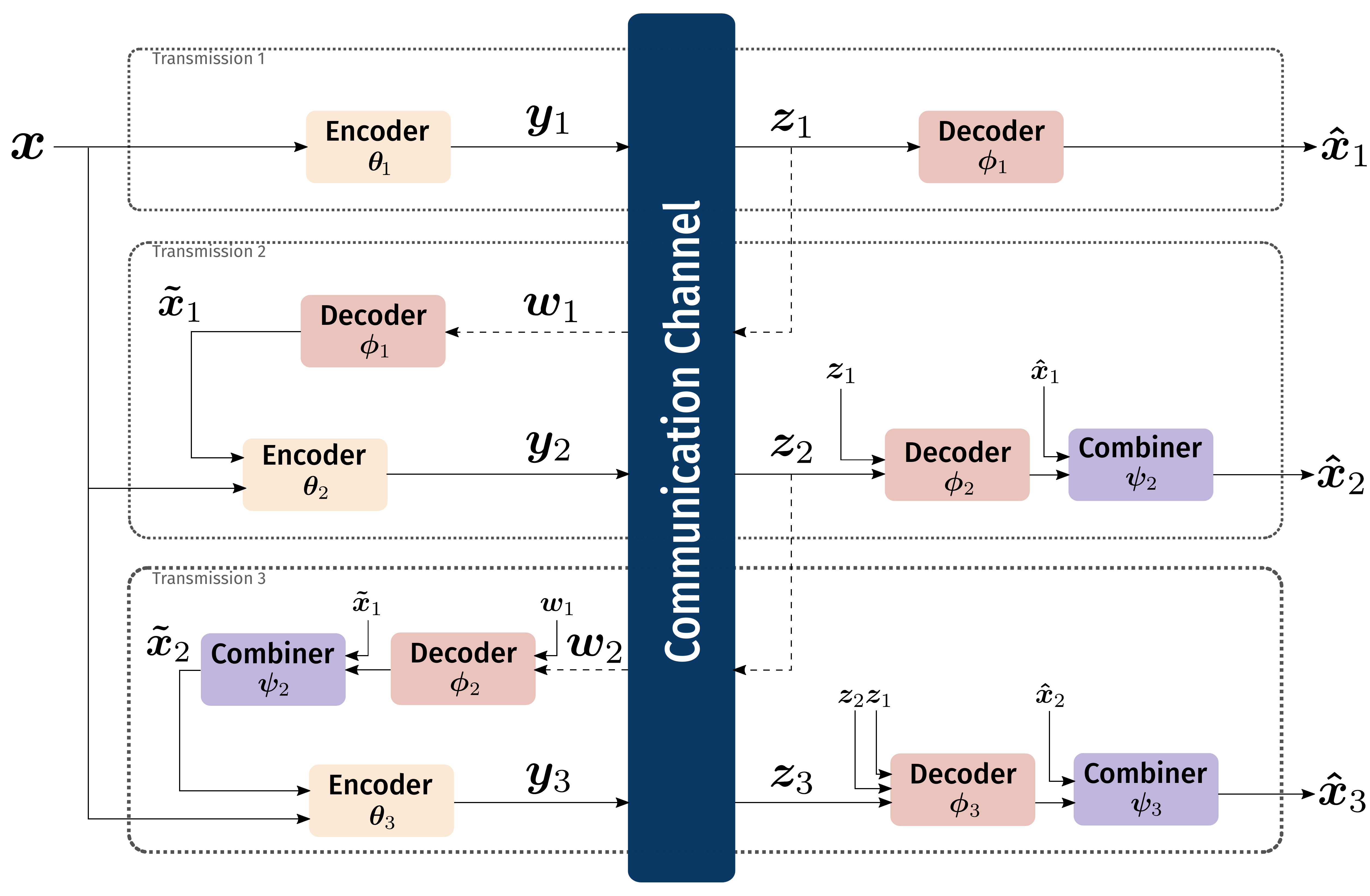}
    \caption{Communication scheme used in this work, detailing individual blocks and inputs/outputs for the case of three partial transmissions ($L=3$).}
    \label{fig:arch_feed}
\end{figure}

\begin{figure}
    \centering
    \includegraphics[width=\linewidth]{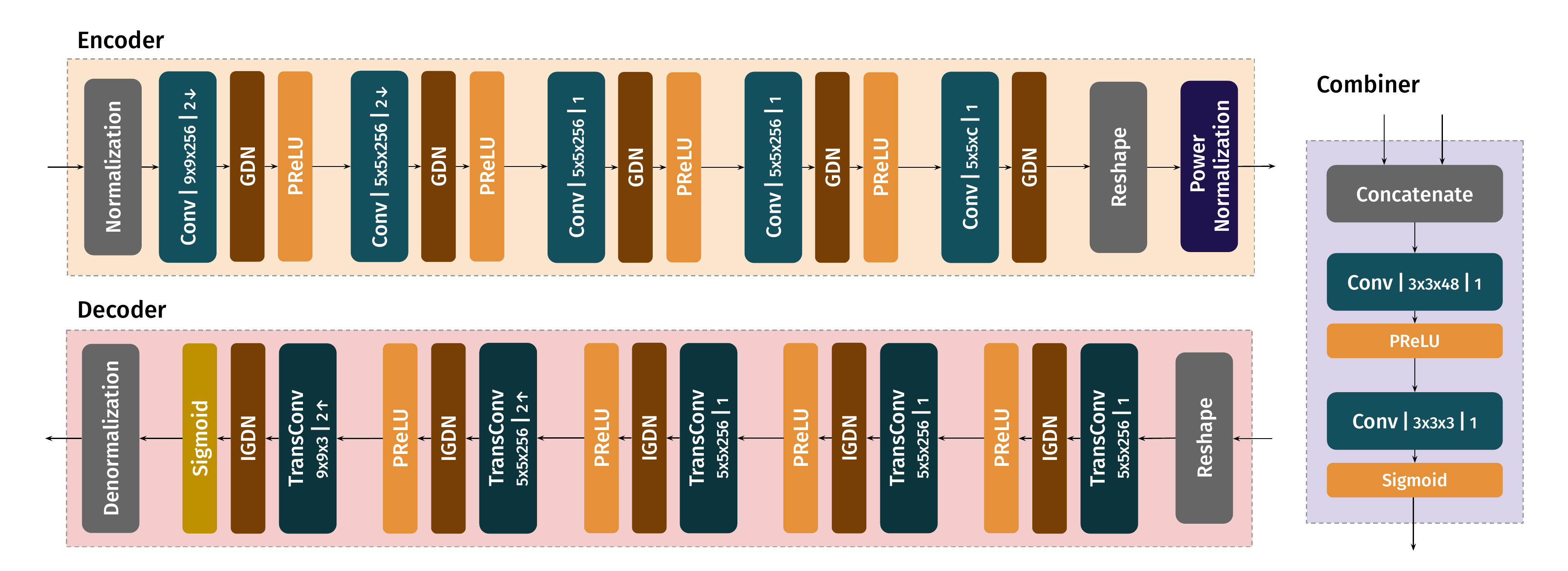}
    \caption{Architecture of the individual blocks from Figure~\ref{fig:arch_feed}, with hyperparameters defined. Convolutions are parameterized by $h \times w \times o ~ | ~ s \updownarrow$, where $h$ and $w$ are the filter height and width, respectively, $o$ the number of channel outputs, and $s$ the stride, which can be either upsampling ($\uparrow$) or downsampling ($\downarrow$) strides.}
    \label{fig:arch_feed_components}
\end{figure}

In this section we will introduce our proposed DNN-based coding scheme that can exploit the channel output feedback, called \mbox{DeepJSCC-$f$}. In terms of architecture, a key innovation of this work is the employment of layered autoencoders, which allows us to take advantage of the available feedback. The proposed design, as seen in Figure~\ref{fig:arch_feed}, models each layer $j$, $j=1, \ldots, L$, as a convolutional autoencoder that is trained for end-to-end communication of images through JSCC.

Motivated by the well-known Schalkwijk-Kailath scheme \cite{Schalkwijk:IT:67, Kailath:PIEEE:67} for the transmission of a single Gaussian source sample over several channel uses, we divide the transmission of each image $\bm x$ into $L$ layers, where each layer tries to improve the quality of the receiver's estimation by transmitting additional information about the residual error. Let layer $j$ be allocated $k_j$ channel uses, where $\sum_{j=1}^L k_j = k$. For the transmission of layer $j$, a channel input vector of ${\bm y}_j \in \mathbb{C}^{k_j}$ is transmitted over the forward channel, resulting in the channel output vector $\bm z_j \in \mathbb{C}^{k_j}$. The noisy feedback vector corresponding to channel output $\bm z_j$ is denoted by $\bm w_j \in \mathbb{C}^{k_j}$. For the encoding of the $j$-th layer, $j > 1$, the encoder has access to the original input $\bm x$ and the channel output feedback of the previous $j-1$ layers, $[\bm{{w}}_1, \dots, \bm{w}_{j-1}]$.  

Note that, when $L=1$ this scheme does not utilize the feedback signal at all. When $L=2$ the encoder benefits only from the feedback signal during the second half of the channel block. Intuitively, one expects the performance to improve as $L$ increases as we would benefit from more and more feedback. On the other hand, as the blocklength of each layer gets smaller, it becomes harder to optimize the corresponding encoder. We will later study the impact of $L$ on the performance of the proposed DNN-based transmission scheme.

Each layer of the proposed \mbox{DeepJSCC-$f$} architecture consists of the following components: (a) an encoder, (b) a decoder, and (c) a combiner, whose functions will be explained below. Each layer contains a copy of all these components (apart from the combiner, that is only used for layers $j \geq 2$). Please see Figure~\ref{fig:arch_feed} for a diagram illustrating the \mbox{DeepJSCC-$f$} architecture for $L=3$ layers.

The encoder at layer $j$ is a deterministic encoding function responsible for producing channel input vector $\bm{y}_j$, $j=1, \ldots, L$. 
In \mbox{DeepJSCC-$f$}, the encoder of each layer $f^{\boldsymbol{\theta}_j}_j$ is modeled as a CNN parameterized by vector $\boldsymbol{\theta}_j$. In the first layer ($j=1$), it receives as input only the source image $\bm{x}$, i.e., we have $\bm{y}_j = f^{\boldsymbol{\theta}_j}_j({\bm x})$. The subsequent encoders $f^{\boldsymbol{\theta}_j}_j$, $j \in \{2, \dots, L\}$, use as input not only the original image $\bm x$, but also an estimate of the image reconstructed by the receiver at the previous layer, $\bm{\tilde{x}}_{j-1}$. We will explain how this estimate is generated below. Accordingly the channel inputs for layers $j=2, \ldots, L$ are given by  $\bm{y}_j = f^{\boldsymbol{\theta}_j}_j({\bm x},\bm{\tilde{x}}_{j-1})$.

The receiver employs a decoder at each layer, which uses all the channel outputs received so far. Similarly to the encoder, the decoder of the $j$-th layer, $g_j^{\boldsymbol{\phi}_j}$, is a CNN parameterized by vector $\boldsymbol{\phi}_j$, $j=1, \ldots, L$, where $\bm{\hat{u}}_j = g_j^{\boldsymbol{\phi}_j}(\bm z_1, \ldots, \bm z_j)$.
After the decoder, for $j \geq 2$, we employ a combiner network $c_j^{\boldsymbol{\psi}_j}$, which recursively combines the output of the decoder of the current layer and the output of the combiner network of the previous layer, that is, we have $\bm{\hat{x}}_{j} = c_j^{\boldsymbol{\psi}_j}(\bm{\hat{x}}_{j-1}, \bm{\hat{u}}_j)$, where $\bm{\hat{x}}_1 = \bm{\hat{u}}_{1}$.

\begin{remark}
We remark here that, in the fixed-length code formulation introduced in Section \ref{sec:ProblemFormulation} we could employ a single combiner network that combines all the decoder outputs to reconstruct the final output $\bm{\hat{x}}$. However, having the intermediate estimates at the end of each layer simplifies the training operation as it allows training the layers sequentially. It also allows us to implement a variable-length coding scheme, in which the encoder can stop transmitting additional layers once the receiver reaches the target reconstruction quality. 
\end{remark}

Note that, we assume that the decoder and combiner parameters $\phi_{j}$ and $\psi_{j}$, $j=1, \ldots, L$, are known both at the receiver and the transmitter.
Thus, the estimation at the transmitter at the $j$-th layer, $\bm{\tilde{x}}_{j-1}$, $j=2, \ldots, L$, is obtained by the transmitter using the received feedback signal $\bm{w}_{j-1}$ at the transmitter, and applying the previous layer's decoder and combiner functions on it, such that $\bm{\tilde{x}}_{j-1} = c_j^{\boldsymbol{\psi}_{j-1}}(\bm{\tilde{x}}_{j-2}, g_{j-1}^{\phi_{j-1}}(\bm{w}_{j-1}))$ for $j>2$, and $\bm{\tilde{x}}_{1} = g_{1}^{\phi_{1}}(\bm{w}_{1})$.
As both the source image $\bm x$ and the estimates at the transmitter $\bm{\tilde{x}}_{j-1}$ have the same dimensions, both inputs are concatenated on the channel axis (thus preserving the height and width positions), and sent as input to the encoder of the next layer $f^{\theta}_j : \mathbb{R}^{2n} \rightarrow \mathbb{C}^{k_j}$, $j=2, \ldots, L$.

\begin{remark}
We remark that, as the encoder performs multiple transmissions, it might not transmit the whole image at every layer, but may instead transmit error correction or refinement information. Note that, instead of imposing this in the architecture, e.g., by feeding only the residual error between the original image and the estimate of the reconstruction at the transmitter to the encoding function, we provide both inputs to the encoder, and let it learn the right encoding function. We have observed that this structure provides significant improvement in the performance. 
\end{remark}

We assume that the forward and the feedback channels are independent of each other, and the transmission of ${\bm y}_j$ sequences is done through independent realizations of a noisy communication channel. Both the forward and the feedback channels are modelled as non-trainable layers (see \cite{Eirina:TCCN:19} for details).
We consider in this work two commonly used channel models: (a) the AWGN channel, and (b) the slow fading channel.
In the case of slow fading channel, we adopt the commonly used Rayleigh slow fading model, where we assume that the channel gain $h$ at each transmission is generated from a circularly symmetric complex Gaussian distribution with zero mean and variance $H_c$.
As all the considered channel models are differentiable, they can be integrated as part of the complete model, without compromising the back-propagation step.

The specific architecture of each component is given by Figure~\ref{fig:arch_feed_components}. It consists of CNN layers, followed by normalization obtained by the generalized normalization transformations (GDN/IGDN)~\cite{balle2015density}, followed by a parametric ReLU (PReLU)~\cite{PReLU} activation function (or a sigmoid, in the last blocks of decoder and combiner). The intuition behind the chosen architecture is that convolutional layers are able to extract the image features, the GDN apply local divisive normalization, which has been shown to be suitable for density modeling and image compression~\cite{Balle:ICLR:17}, while the non-linear activations allow the learning of non-linear mapping from the source signal space to the channel input space, and vice versa.  The exact hyper-parameters for each block, shown in Figure~\ref{fig:arch_feed_components}, were chosen based on experimentation and inspired by~\cite{BalleICLR2018} and~\cite{Eirina:TCCN:19}.

The parameter $c$ in the encoder's last CNN layer is responsible for defining the dimension of the channel input $k_j$. Considering an image input with dimensions $n = H \times W \times C$, where $H$ is the image height, $W$ its width and $C$ the number of channels ($3$ for colored images, $1$ for grayscale), and the dimension reductions caused by downsampling, the output of the enconder's last CNN block has dimension  $2k_j = H/4 \times W/4 \times c$. This tensor is then reshaped as a one dimensional complex latent vector $\bm{\tilde y}$ of dimension $1 \times k$, to be transmitted over the channel.  Before transmission, the latent vector is normalized to enforce the average power constraint and input images are normalized by the maximum pixel value $255$, producing values in the $[0,1]$ range. This operation is reverted at the decoder, before output. We note that imposing the average power constraint at each layer results in a stricter constraint compared to the one required by the model, which allows arbitrary transmit power over any sub-block as long as the average power constraint is satisfied over the whole blocklength of $k$ channel uses. We will explore adaptive power allocation strategies through reinforcement learning as a future extension of this work, which we believe can provide further improvements, especially if the channel state varies over the transmission of a single image.

The model is trained gradually, layer by layer. Each layer attempts to minimize the average distance between the input image $\bm{x}$ and its partial reconstruction $\bm{\hat x}_j$:
\begin{equation}
    (\theta_j^*, \phi_j^*, \psi_j^*) = \underset{\theta, \phi, \psi}{\mathrm{arg\,min}}~ \mathbb{E}_{p(\bm{x},\bm{\hat{x}}_j)} [d(\bm{x}, \bm{\hat{x}}_j)],
    \label{eq:expdistortion}
\end{equation}
where $d(\bm{x}, \bm{\hat{x}}_j)$ is a given distortion measure, and $p(\bm{x},\bm{\hat{x}}_j)$ the joint probability distribution of the original and reconstructed images.
Upon convergence, the layer parameters are fixed and additional layers are trained, using the previous channel output as feedback information. Note that, although previous encoders and decoders have fixed weights, every channel realisation is independent and produce new random variables.
As mentioned above, this gradual approach not only simplifies training, but by producing a partial reconstruction $\bm{\hat{x}}_j$ of the complete input image at each layer, it also enables variable-length coding by allowing to stop transmission when a target quality level is met. 

\section{Experimental Results}

In this section we present a set of experiments evaluating the performance of \mbox{DeepJSCC-$f$} in various different scenarios, and comparing it with several benchmarks. In particular, we will consider both perfect (noiseless) and noisy channel output feedback, different compression rates, both AWGN and fading channels at different channel SNR values. We should remark that, to the best of our knowledge, \mbox{DeepJSCC-$f$} is the first practical JSCC implementation in the literature able to transmit images exploiting channel output feedback, so there are no direct competitor to the results presented here. We will consider practical schemes that ignore the feedback signal, or compare our results with theoretical bounds, in particular assuming capacity-achieving channel codes.

Unless stated otherwise, all results from this section were obtained by models trained and evaluated using images with dimension $32 \times 32 \times 3$ (height, width, channels) from the CIFAR10 dataset~\cite{CIFARdataset}. The training and evaluation datasets are composed of completely distinct images and contain $50000$ and $10000$ images, respectively.
All plotted results are average values obtained from $10$ realizations of the channel for every image on the evaluation dataset.

The model was implemented\footnote{The code is available in \url{www.github.com/kurka/deepJSCC-feedback}} in Tensorflow~\cite{abadi2016tensorflow} and optimized using the Adam algorithm~\cite{AdamICLR2015}. We used a learning rate of $10^{-4}$ and a batch size of $128$. Models were trained until the performance of a validation set (selected and separated from the training dataset) stop decreasing.

The loss function used for the training of the model at layer $j$ is given by the average mean squared error (MSE) over $N$ image samples:

\begin{equation}
\mathcal{L} = \frac{1}{N} \sum_{i=1}^N ||\bm x^i -\hat{\bm x}_j^i ||^2,
\label{eq:mseloss}
\end{equation}
where, with abuse of notation, ${\bm x}^i$ denotes the $i$-th image sample in the dataset, and $\hat{\bm x}_j^i$ its reconstruction at layer $j$. 

In order to measure the performance of our algorithm and alternative schemes, we use the peak signal-to-noise ratio (PSNR), given by:

\begin{equation*}
  \mathrm{PSNR} = 10\log_{10}\frac{\mathrm{MAX}^2}{||\bm x-\hat{\bm x}_j ||^2} ~ (dB),
\end{equation*}
where $\mathrm{MAX}$ is the maximum value a pixel can take, which is $255$ as we use 24-bit depth RGB images.

To measure the quality of a channel in which communication is performed, we consider the average signal-to-noise ratio (SNR) given by:

\begin{equation*}
    \SNR = 10\log_{10}\frac{1}{\sigma^2} ~ (dB),
\end{equation*}
representing the ratio of the average power of the coded signal (channel input signal) to the average noise power. 

\subsection{\mbox{DeepJSCC-$f$} with  Two Layers ($L=2$)}

We first consider the case in which the feedback channel is used only once per transmission and perfect channel output knowledge is assumed. Thus, the source image $\bm x$ is transmitted in two layers: first a base layer $\bm{y}_1$ with bandwidth $k_1$ is sent over the channel; then, using the channel output corresponding to $\bm{y}_1$, $\bm{w_{1}} = \bm{z}_1$, a second message $\bm{y}_2$, of length $k_2$ is transmitted. For simplicity, we set $k_1 = k_2$.

\subsubsection{General Performance}

\begin{figure}
    \centering
    \includegraphics[width=\linewidth]{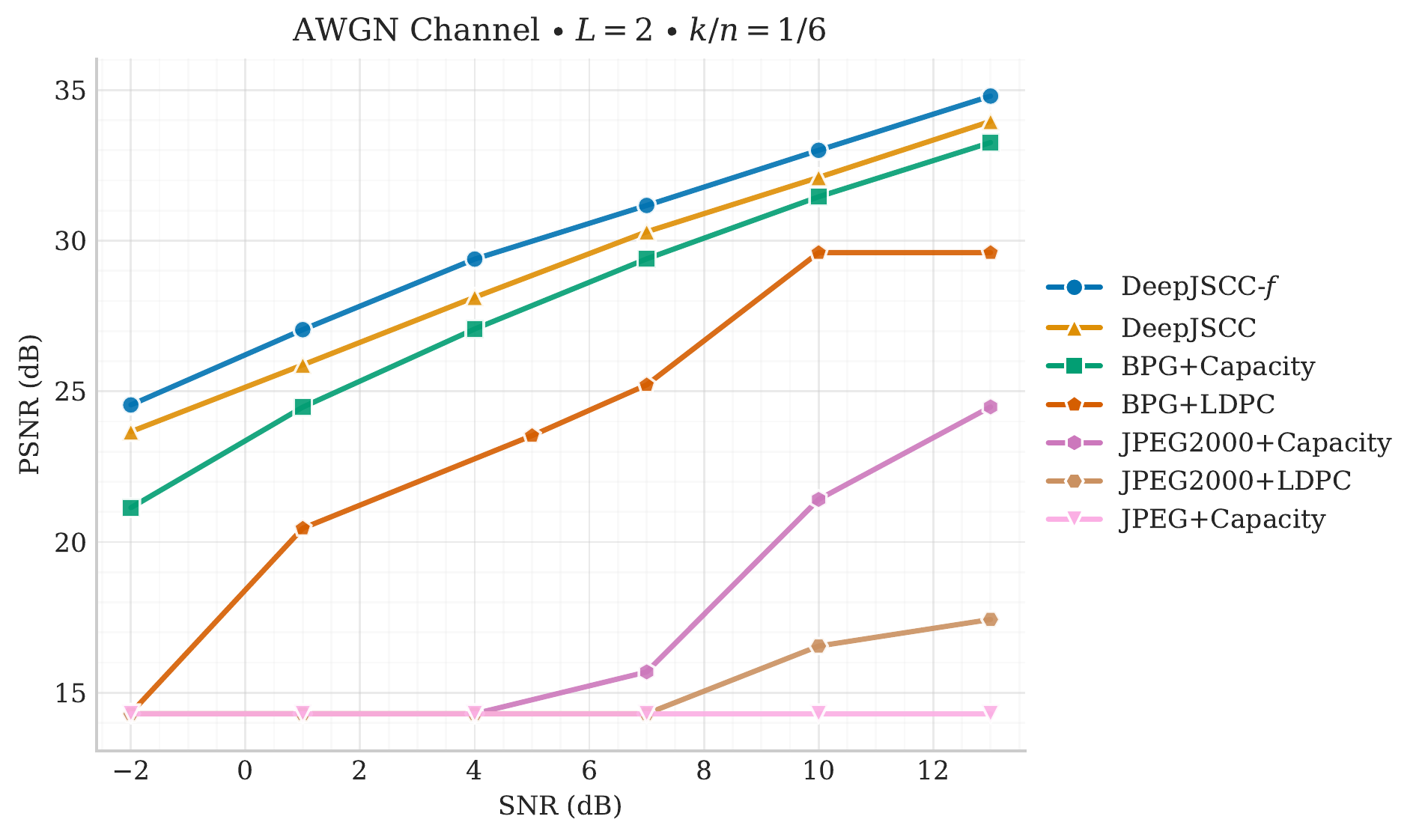}
    \caption{Performance comparison as a function of the channel SNR, demonstrating the gains of using feedback information, compared to a single transmission scheme and a number of separation-based digital schemes.}
    \label{fig:snr_1fb}
\end{figure}

Figure~\ref{fig:snr_1fb} shows the results for compression rate $k/n = 1/6$, performed in two stages with $k_1/n = k_2/n = 1/12$. The forward channel is affected by AWGN, and we consider noiseless feedback. The performance of models subject to different forward channel SNRs is presented in the plot, where a separate set of encoders, decoders, and combiners is trained for each SNR.

We first compare our scheme to the state of the art JSCC scheme for image transmissions, without the use of feedback. For this, we use the DeepJSCC algorithm from~\cite{Eirina:TCCN:19}, further enhancing it by employing the architecture shown in Figure~\ref{fig:arch_feed_components}. This is also the special case of our scheme with $L=1$; that is, a single transmission with a bandwidth ratio $k/n = 1/6$. We see that \mbox{DeepJSCC-$f$} brings a considerable performance improvement, outperforming the scheme without feedback by a margin of at least 1dB, being particularly more efficient in the low SNR regime, when the feedback information is more relevant. This superior performance can be explained solely by the use of feedback, given that the components share the same architecture. Note also that, sending a source over multiple layers without feedback can at best maintain, or decrease the performance, compared to single-layer transmission~\cite{Kurka:ICASSP19}.

We then compare the performance with separation-based digital schemes, where the images are first compressed using state of the art compression codecs, and then encoded by a channel code against channel noise or fading. Note that, in the separate architecture feedback can only be used to improve the channel coding performance. For the source code, we consider well established image compression algorithms: JPEG, JPEG2000\footnote{\url{https://jpeg.org/jpeg2000/index.html}} and BPG\footnote{\url{https://bellard.org/bpg/}}. For fair comparison, we remove the header information from the compressed files, so we are only considering the communication of the compressed bits. These codecs are the fruit of years of research and development, and are widely used in diverse applications, such as content delivery over the Internet, digital photography and medical imaging. Although there has been recent advances in DNN-based image compression methods, the state of the art still performs (in terms of PSNR) similarly to BPG on the CIFAR10 dataset~\cite{Minnen:NIPS:18}. Therefore, we chose not to include it in this comparison.
For the channel code, we consider a practical scheme -- low-density parity-check code (LDPC) followed by quadrature amplitude modulation (QAM) -- and a theoretical bound -- channel capacity. For the LDPC+QAM scheme, we evaluate the performance of different combinations of code rates and modulations, presenting here the envelope of the best performing configurations for each forward channel SNR (more details are given in Appendix~\ref{sec:ldpc}). The capacity achieving code would be a hypothetical channel code, achieving the same rate as the underlying channel capacity. This is not achievable in practice, especially in short blocklengths (we have $k=512$ for the setting considered in Figure~\ref{fig:snr_1fb}, as the bound assumes infinite blocklength ($k \to \infty$). To evaluate the corresponding performance bound, we find the compression rate dictated by the channel capacity and the bandwidth ratio, and evaluate the PSNR obtained by the employed compression codec for each image at that bit rate. The resulting performance serves as an upper bound on the performance of any separation-based transmission scheme with feedback (using the particular compression scheme), as the channel capacity does not increase with feedback.
If, for a given image, bit rate is below the minimum feasible bit rate, we consider that the receiver reconstruct, for each color channel, the mean value of all the pixels for that channel.

We plot in Figure~\ref{fig:snr_1fb} the performance of different combinations of source and channel codes. We see that JPEG, currently the most popular and the most widely employed image compression codec, presents the worst performance, not being able to compress with enough quality in low SNRs, resulting in failed transmissions; while BPG is the best performing algorithm. We also note that our algorithm can surpass, particularly in the low SNR regime, BPG combined with a capacity-achieving channel code. Since BPG performs similarly or better than DNN-based image compression schemes for the considered image dataset, the improvement here can be attributed to the improvement from JSCC. We should remark that we have been quite generous to the separation-based scheme by considering capacity achieving channel codes despite operating in a short blocklength regime. Hence, the real improvement with respect to practical codes will be  even more significant.

Hyeji et al.~\cite{Hyeji:NIPS:2018} present a practical DNN-based channel code, Deepcode, that exploit the feedback channel, outperforming other codes such as LDPC and Turbo code. Although we have considered Deepcode as channel code for our experiments, we found that the proposed scheme cannot be easily employed for the transmission of images in large blocklengths. The results presented in \cite{Hyeji:NIPS:2018} consider transmission of a message of 50 bits at rate $1/3$ on the forward channel SNR range -2dB -- 1dB. In this range (-2dB -- 1dB), indeed Deepcode outperforms other channel codes; however, we found that none of our source codes could achieve a non-trivial PSNR value that improves upon the quality obtained when averaging all pixels with their average values (lower bound considered in case of failed transmissions). Therefore, in low SNR regimes, the separation-scheme cannot benefit from the more capable channel code of \cite{Hyeji:NIPS:2018}. However, if the forward channel improves ($\SNR > 2dB$), we noticed that LDPC outperforms the Deepcode, not justifying its use. Therefore, we decided not to plot those results in Figure~\ref{fig:snr_1fb}, but highlight again that their performance would still be below the one obtained by assuming a capacity-achieving channel code.

\subsubsection{Graceful Degradation}\label{ss:graceful_noiseless}

\begin{figure}
    \centering
    \includegraphics[width=.7\linewidth]{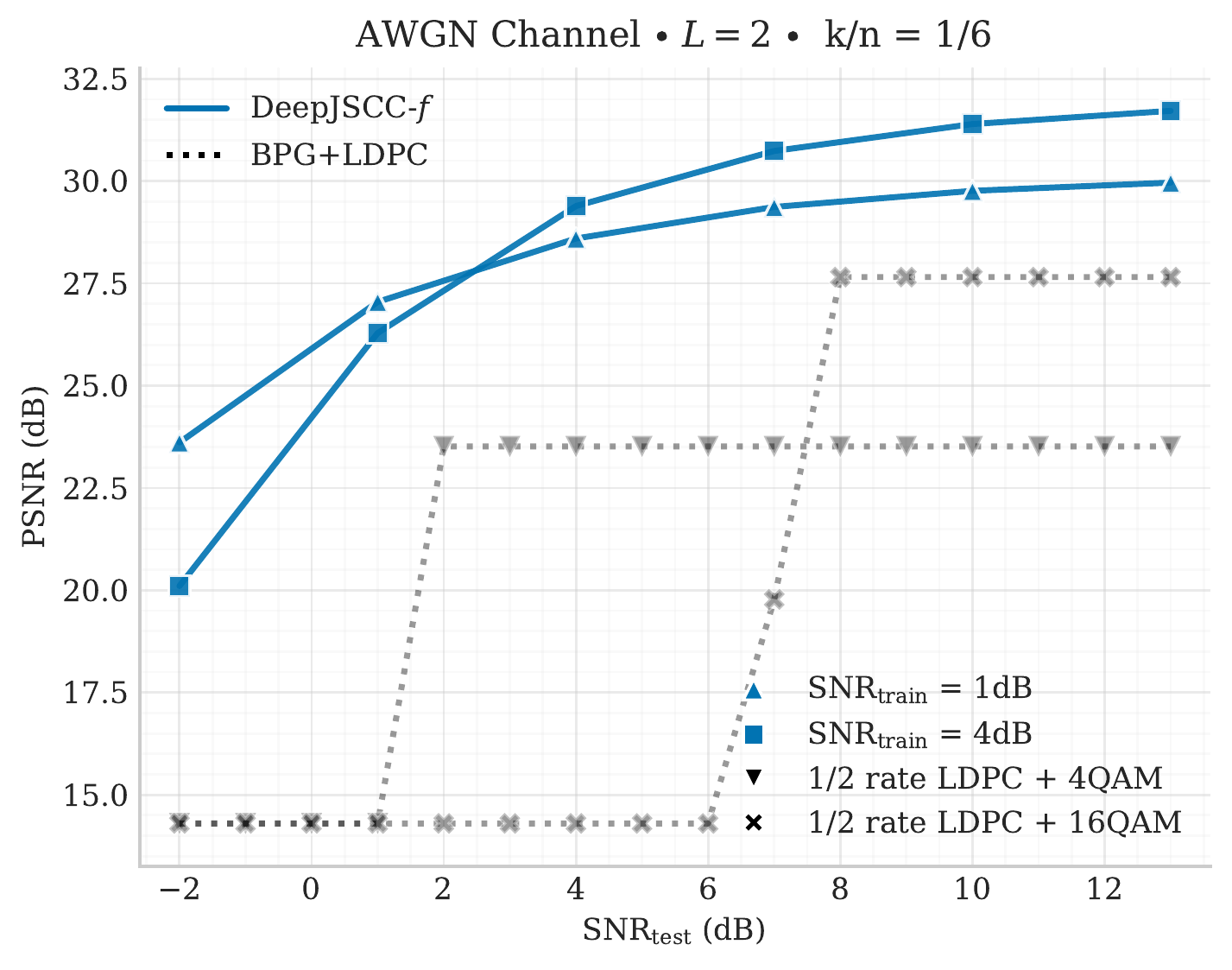}
    \caption{Visualization of the graceful degradation achieved by \mbox{DeepJSCC-$f$} (analog behaviour) versus the cliff effect of separation-based (digital) schemes when there is discrepancy between the channel conditions during design ($\SNR_\mathrm{train}$) and deployment ($\SNR_\mathrm{test}$).}
    \label{fig:graceful}
\end{figure}

In addition to improving the performance with respect to separation-based transmission, another advantage of DeepJSCC with respect to digital schemes is the graceful degradation it exhibits with the channel SNR~\cite{Eirina:TCCN:19}. Figure~\ref{fig:graceful} shows the performance when two models trained with $\SNR_\mathrm{train} = 1dB$ and $\SNR_\mathrm{train} = 7dB$ (colored curves) are evaluated in a range of different $\SNR_\mathrm{test}$ values, in the range of -2dB to 13dB. Note how the performance increases when $\SNR_\mathrm{test} > \SNR_\mathrm{train}$; and, although the performance degrades as $\SNR_\mathrm{test}$ decreases, it does not drop rapidly, and it is still possible to retrieve recognisable images, even at very low SNRs. The digital scheme, in comparison, does not improve when the channel quality increases, and if the channel deteriorates beyond a threshold value, the performance decreases drastically as the receiver cannot decode the channel code, and hence the source code, any more (known as the \textit{cliff effect}).

\subsubsection{Fading Channel}\label{ss:fading}

Here we consider a slow Rayleigh fading channel, whose gain changes randomly according to a Rayleigh distribution, but remains constants during the transmission of a single image. 
Previous work~\cite{Eirina:TCCN:19} has shown that the single-layer DeepJSCC is capable of learning representations for the fading channel with remarkable performance, outperforming digital schemes. The most surprising result is the fact that DeepJSCC is able to learn an efficient transmission scheme without the need of transmission of pilots, or explicit channel estimation, while the separation-based scheme inherently assumes perfect CSI at the receiver.

\begin{figure}
    \centering
    \includegraphics[width=.7\linewidth]{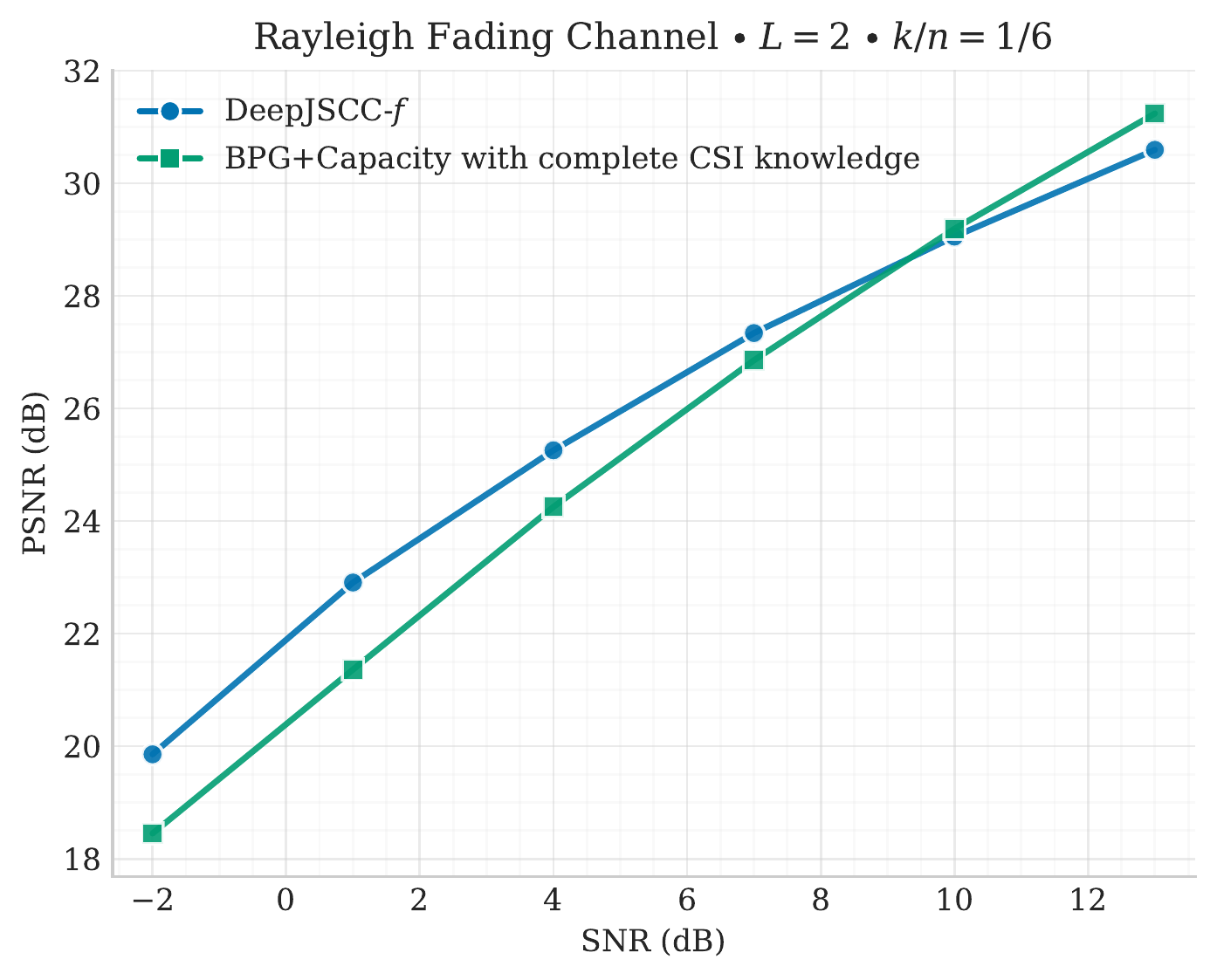}
    \caption{Performance of \mbox{DeepJSCC-$f$} when trained over a slow Rayleigh fading channel with different average SNRs.}
    \label{fig:snr_1fb_fading}
\end{figure}

In Figure~\ref{fig:snr_1fb_fading}, we plot the results for a range of average channel SNRs, for a bandwidth ratio of $k/n=1/6$. We compare the performance of \mbox{DeepJSCC-$f$} with that of a separation-based scheme that again employs BPG followed by the instantaneous capacity of the channel. Note that this inherently assumes that the transmitter (as well as the receiver) has access to the perfect CSI due to the feedback link. It is remarkable that \mbox{DeepJSCC-$f$} can still outperform the digital scheme despite all the impractical assumptions to the advantage of the latter.

\subsection{\mbox{DeepJSCC-$f$} with Multiple Layers ($L>1$)}

\subsubsection{Impact of $L$}

As we have mentioned before, one would expect the performance of the proposed \mbox{DeepJSCC-$f$} scheme to improve with $L$ as the feedback resources are better utilized as $L$ increases. On the other hand, \mbox{DeepJSCC-$f$} is designed to transmit and refine the original image at each layer. Hence, we expect that allocating a channel bandwidth below a certain value will not allow a reasonable quality reconstruction of the image. 

\begin{figure}
    \centering
    \includegraphics[width=.7\linewidth]{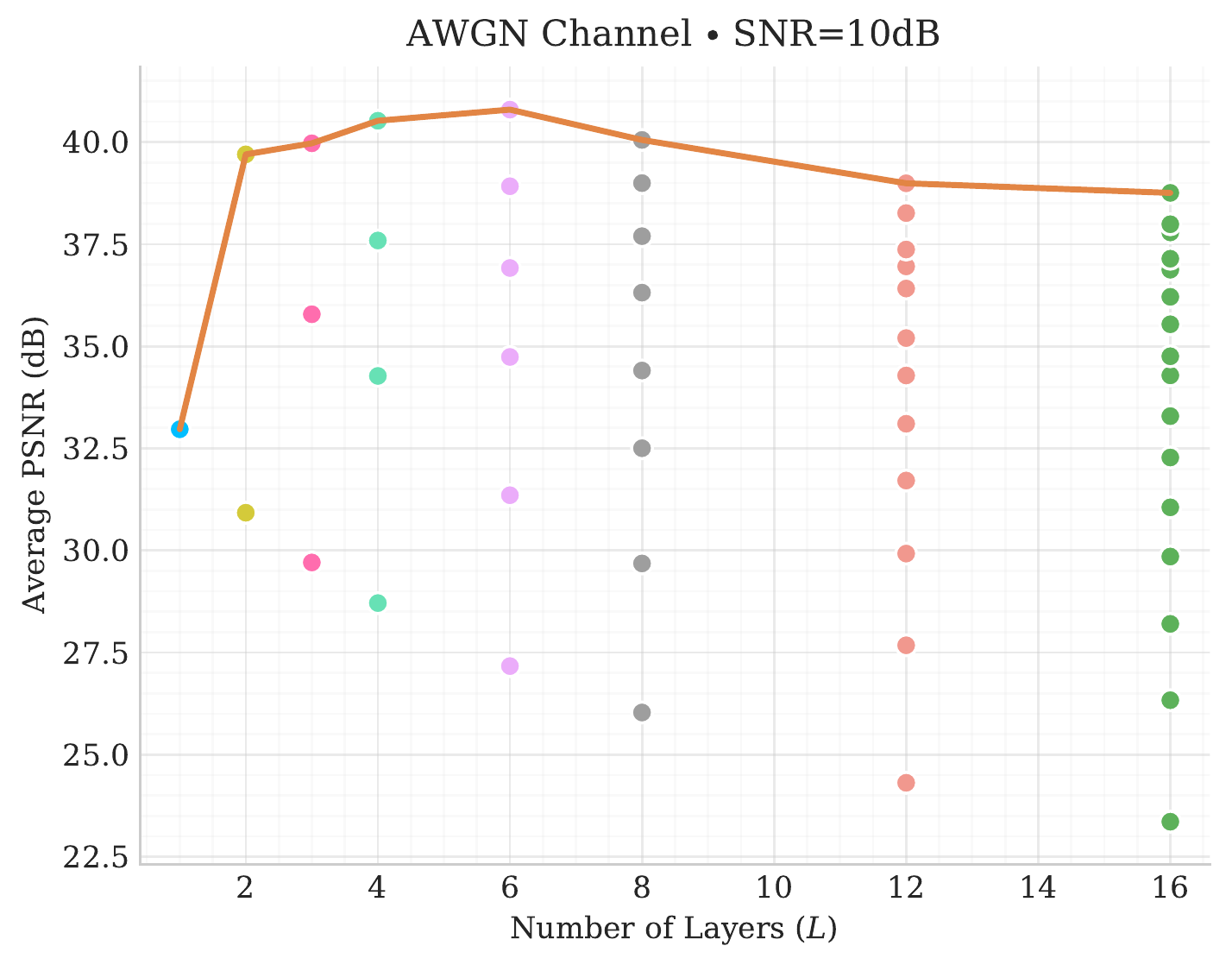}
    \caption{Impact of amount of feedback in the performance. Initially, the addition of feedback rounds increases performance, but after a point, the gains stagnate.}
    \label{fig:layer_psnr}
\end{figure}

We plot the performance achieved for different $L$ values in 
Figure~\ref{fig:layer_psnr} for a fixed bandwidth ratio of $k/n = 1/2$). For each $L$ value, we mark the average PSNR value achieved by each layer, starting from the first to the $L$th. The final performance for each $L$ value is the highest point in the plot. 

We see that the addition of new layers increases the performance initially. However, as more layers are introduced, the performance stabilises and even declines after a certain threshold, as expected. 
Also, increasing $L$ directly increases the complexity of the model, as we need to train a separate set of neural networks (encoder, decoder, and a combiner) for each layer.  Our simulation results suggest that $L=4$ layers typically provides a reasonable performance trade-off.

\begin{figure}
    \centering
    \includegraphics[width=.7\linewidth]{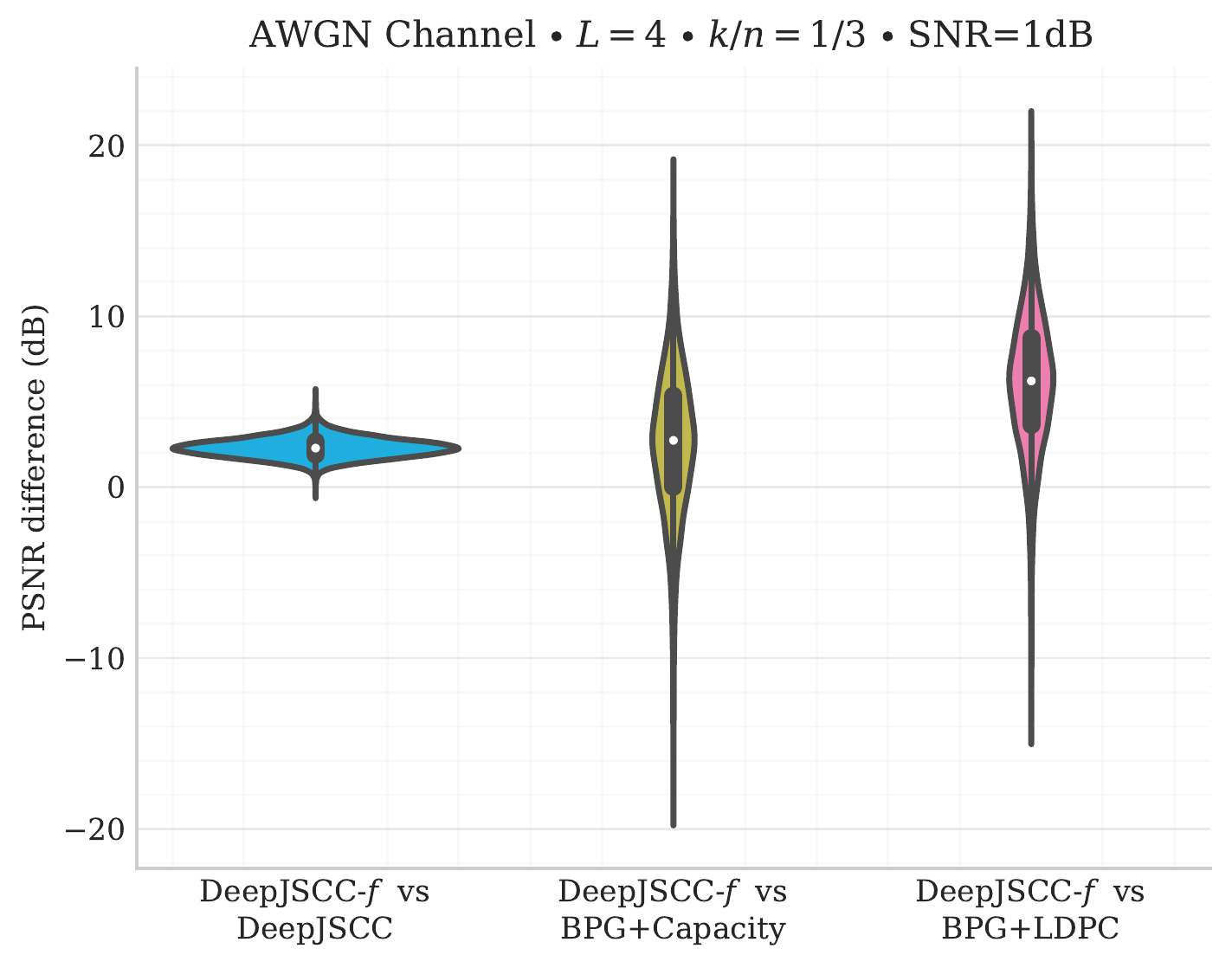}
    \caption{Performance gap distribution between \mbox{DeepJSCC-$f$} and other schemes over the CIFAR-10 dataset. \mbox{DeepJSCC-$f$} produces superior results (positive gaps) for the vast majority of images transmitted.}
    \label{fig:confidence}
\end{figure}

We plot the performance of \mbox{DeepJSCC-$f$} with $L=4$ layers in Figure~\ref{fig:snr4} for a range of channel SNR values and a compression ratio of $k/n = 1/3$. Here we observe that the gain of \mbox{DeepJSCC-$f$} with respect to the separation-based bound is even more pronounced, with gains up to 3~dB in average PSNR in the low channel SNR regime.
We further investigate and verify the superiority of \mbox{DeepJSCC-$f$} by, instead of comparing the average PSNR of each scheme (as in  Figure~\ref{fig:snr4}), comparing the performance \emph{gap} between transmissions at a specific SNR (1dB), using different schemes for all images of the CIFAR10 dataset. Figure~\ref{fig:confidence} shows the distribution of the PSNR difference between \mbox{DeepJSCC-$f$} and other schemes, where positive values represent occurrences in which \mbox{DeepJSCC-$f$} outperforms its competitor. We clearly see that for the vast majority of the images transmitted our model results in an improvement in the reconstruction quality.

\begin{figure}
    \centering
    \includegraphics[width=\linewidth]{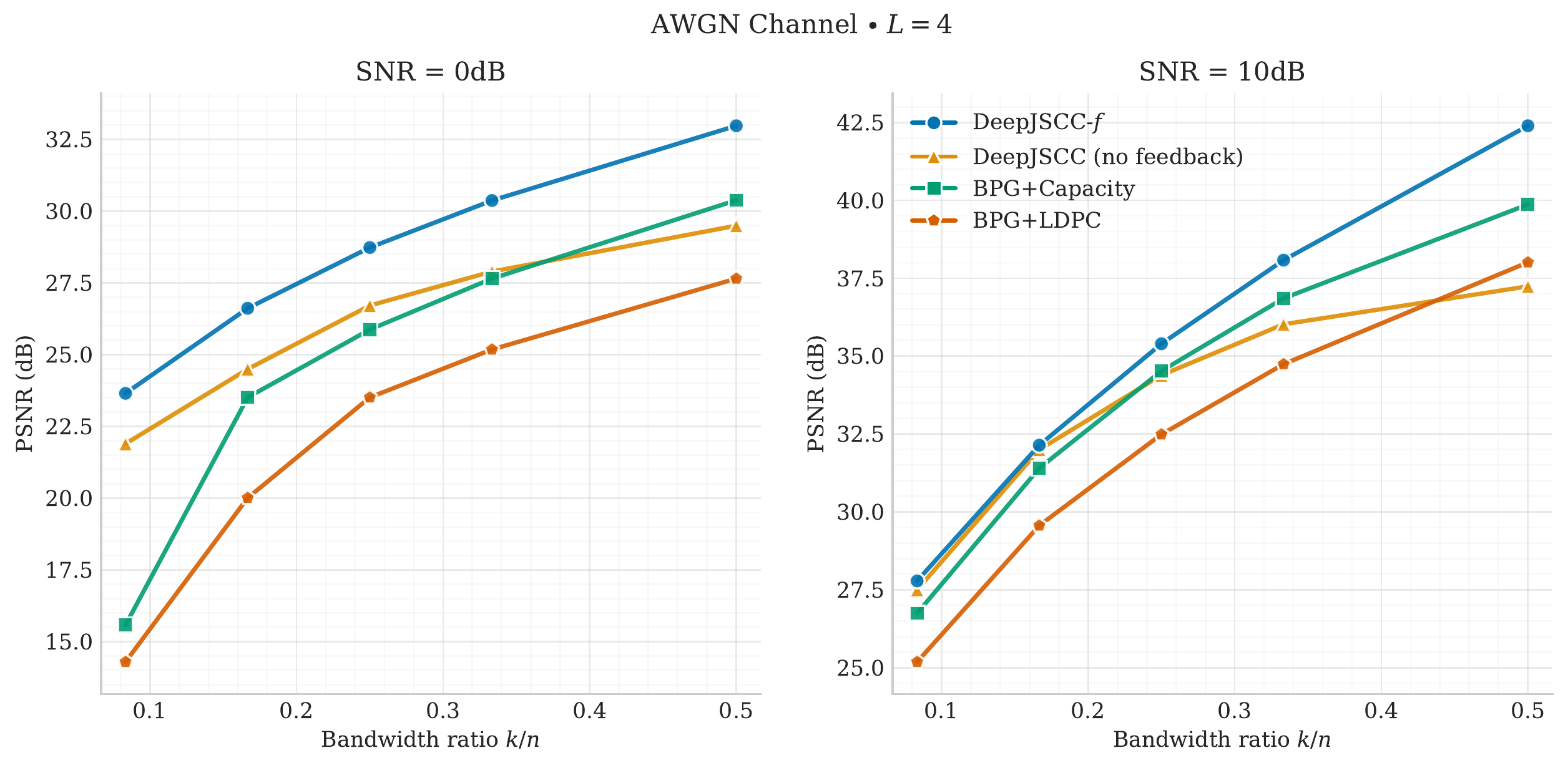}
    \caption{Performance comparison for different bandwidth ratios, for $L=4$.}
    \label{fig:cr_noiseless}
\end{figure}

Finally, in Figure~\ref{fig:cr_noiseless} we present results for a wider range of compression ratios and two different channel SNRs. We observe that DeepJSCC-$f$ outperforms all other benchmarks, and even the separation-based bound (which becomes looser as the bandwidth ratio gets smaller), at all settings.
Thus, our results show the versatile performance of DeepJSCC-$f$ over a wide array of configurations, presenting itself as a very efficient image transmission method.

\subsubsection{Variable Rate Transmission}\label{ss:variable_rate}

We can formulate the JSCC coding problem differently by setting a certain quality target for the delivery of each image, and aim at minimizing the required channel bandwidth. In this case the perfect channel output feedback provides the transmitter with the knowledge of the stopping time. It is shown in \cite{Kostina:IT:17} through theoretical analysis that allowing variable rate coding leads to a significant improvement in the delay-distortion trade-off. 

As we highlighted earlier the layered architecture used in \mbox{DeepJSCC-$f$} naturally lends itself to such a variable transmission scheme. Since the receiver reconstructs the image at the end of the transmission of each layer using a deterministic decoding function, the encoder will know exactly after each layer whether it needs to send further information, or it can stop transmission as the distortion target is met.

We experiment this setting by considering the transmission with $L=8$ layers, and computing the average bandwidth needed for achieving a target PSNR. We compare this to a digital scheme transmitting headerless BPG with an ideal capacity-achieving code. For the digital scheme, we compress each image to the minimum amount of bits that meet the target PSNR value, and find how many channel uses is needed to transmit so many bits over the channel, assuming a capacity-achieving channel code.

\begin{figure}
    \centering
    \subfloat[]{%
        \includegraphics[width=0.49\textwidth]{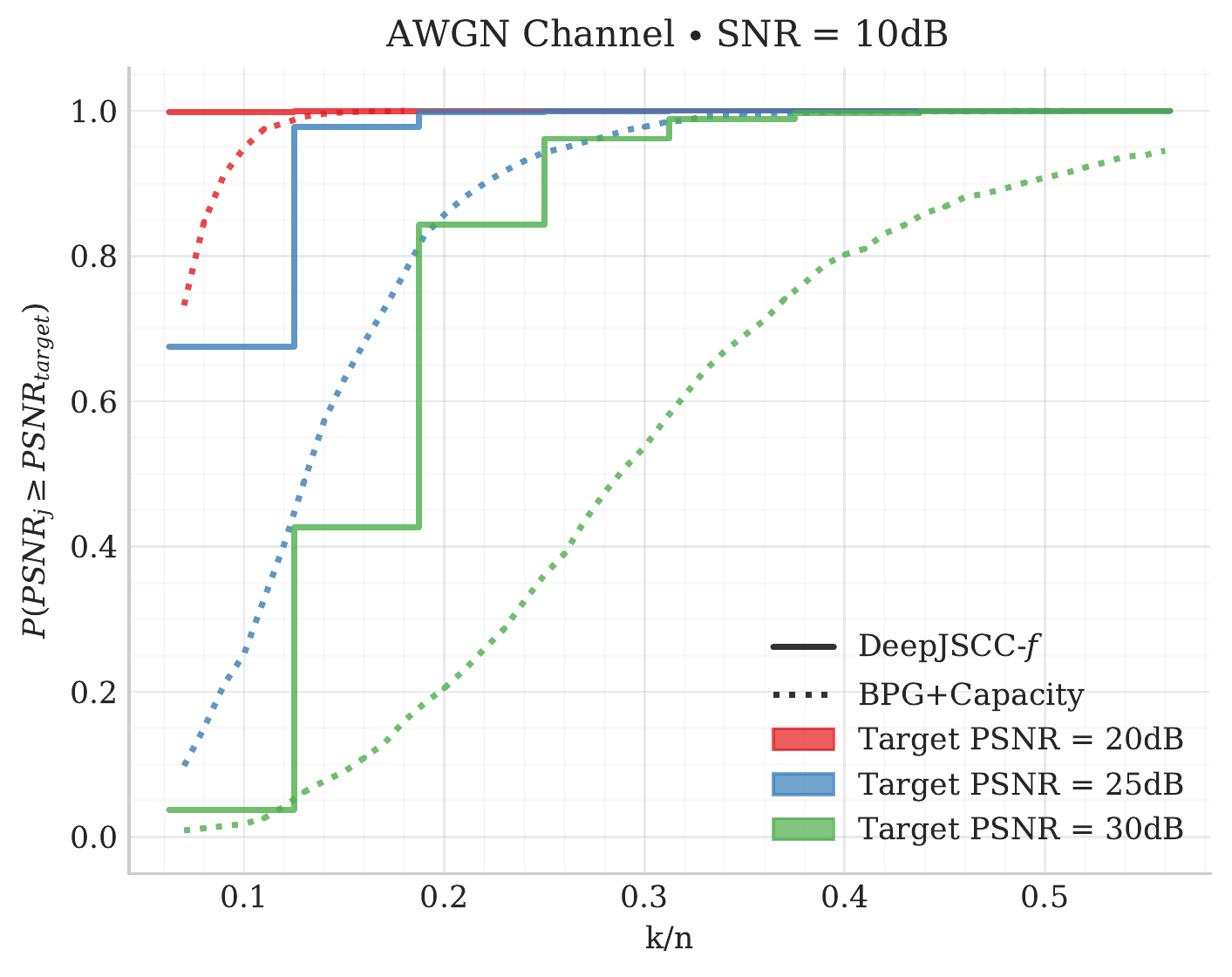}
        \label{fig:targetcdf}}
    \hfill
    \subfloat[]{%
        \includegraphics[width=0.49\textwidth]{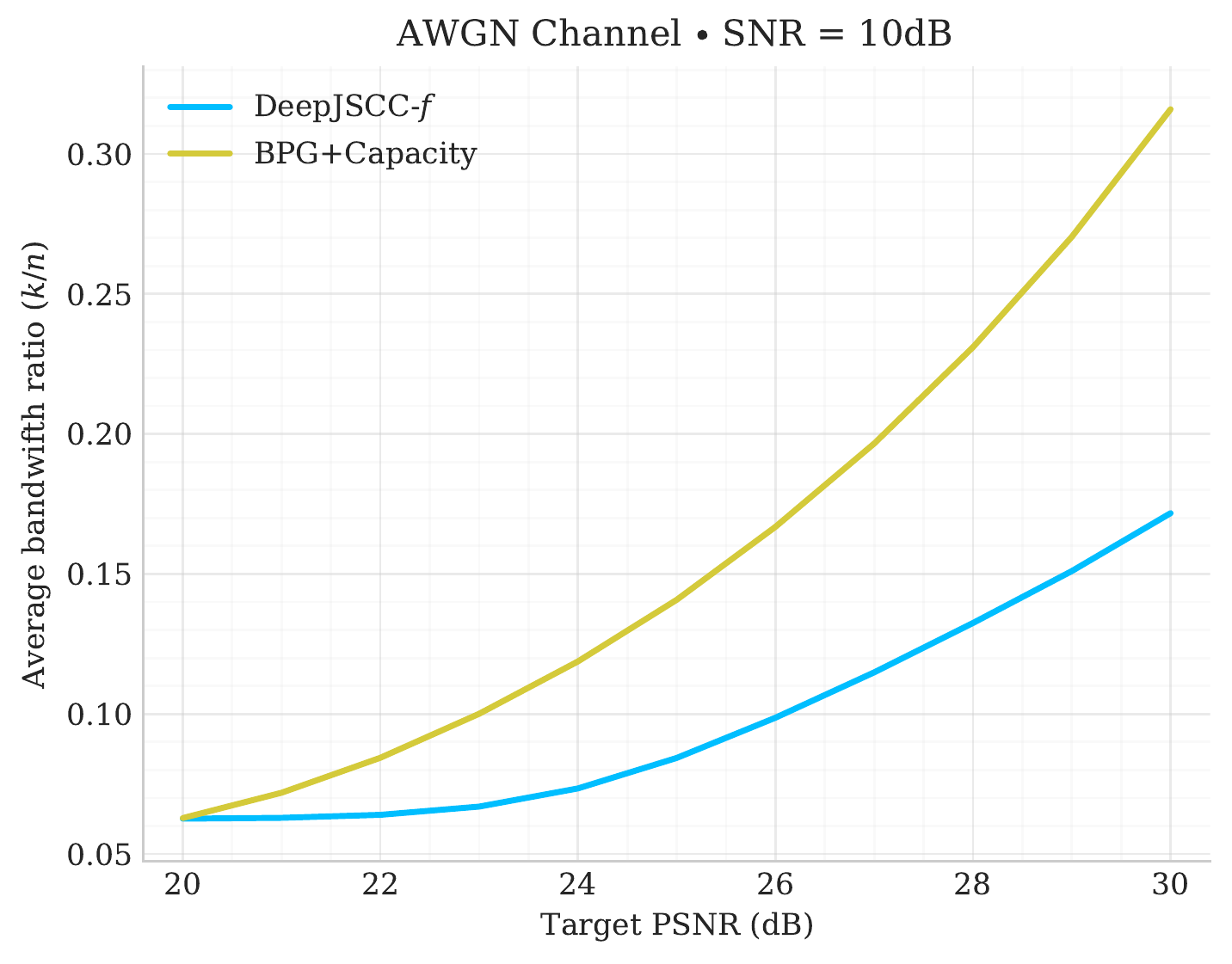}
        \label{fig:target}}
    \caption{Target SNR analysis, presenting (a) the cumulative distribution function of bandwidths given a specific target PSNR for \mbox{DeepJSCC-$f$} over the CIFAR10 training dataset, and (b) the average bandwidth ratio to achieve a specific target PSNR. Note that our scheme outperforms the separation-based scheme in the whole range, especially in higher target PSNRs.}\label{fig:targets}
\end{figure}

In Figure \ref{fig:targetcdf} we plot the cumulative distribution function of the necessary bandwidth for achieving three different PSNRs (20dB, 25dB and 30dB).
We can see that, despite the low granularity of the \mbox{DeepJSCC-$f$} scheme (transmission can be stopped only after transmitting a whole layer), it provides significant improvement compared to any separation-based scheme. Almost all the images require less channel bandwidth resources with \mbox{DeepJSCC-$f$} compared to separation. Note that it is possible to have more granularity by decreasing the number of channel uses employed by each layer, which would provide a more accurate mapping to the desired PSNR level, at the cost of increased complexity.

In Figure~\ref{fig:target} we plot the average bandwidth ratio required to meet different target SNR values. Significant gap between the two curves confirm the theoretical results in this practical setting. We also observe that the gap increases with the PSNR target. For a PSNR target of $30$~dB, the average bandwidth ratio is almost half that is required by the bound on the separation-based schemes.

We should again highlight the fact that this bound is particularly loose when the image can be transmitted with only a few layers as this would correspond to a very short blocklength. For example, as we can see in Figure \ref{fig:targetcdf}, when we target a PSNR value of $20$~dB over a channel with $\SNR = 10$~dB, almost all of the images bits should be transmitted reliably within only $307$ channel uses (i.e., $k/n = 0.1$) according to the bound considered.

\subsubsection{Fading Channel}\label{ss:fading channels}

\begin{figure}
    \centering
    \includegraphics[width=0.7\linewidth]{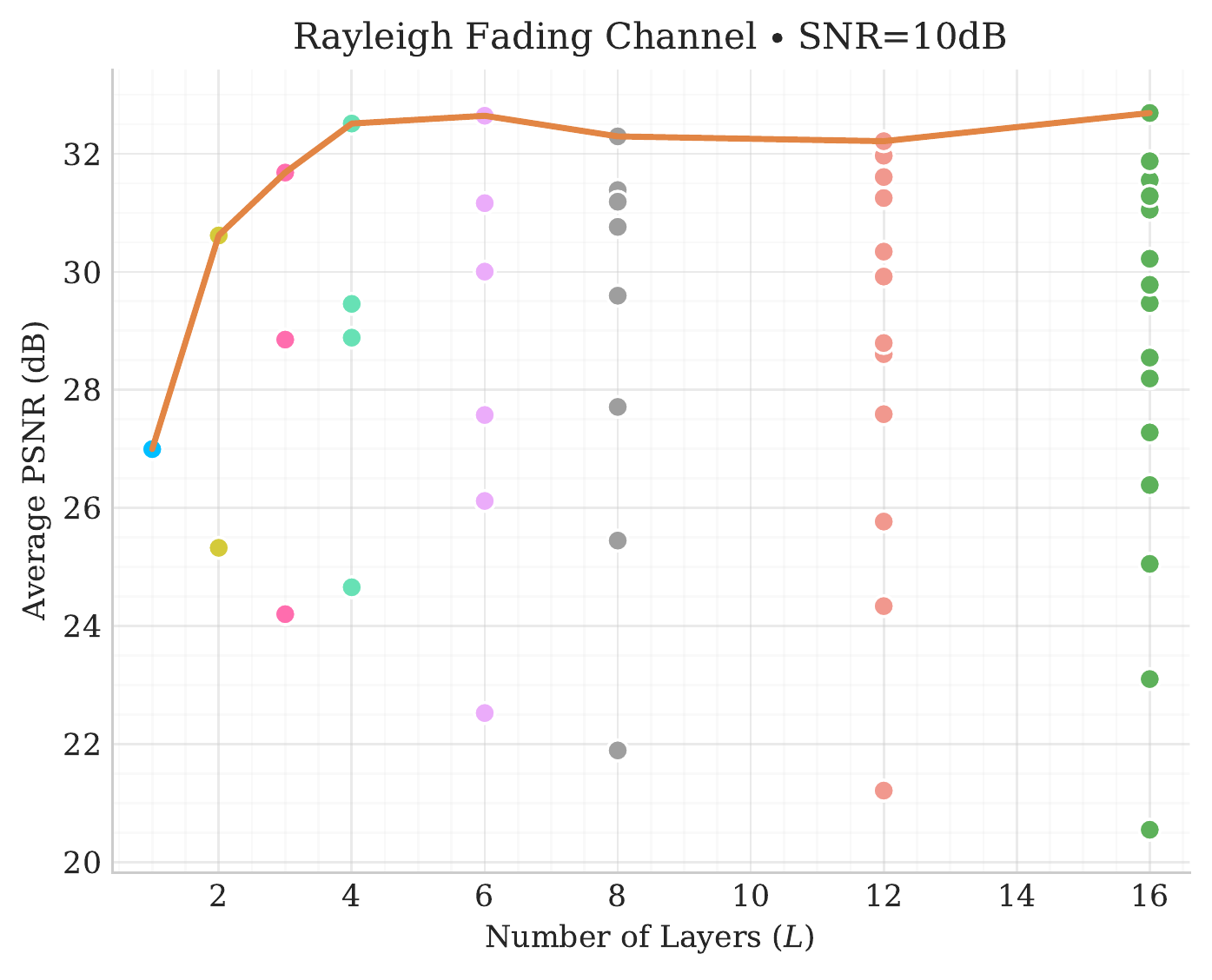}
    \caption{Performance of \mbox{DeepJSCC-$f$} in fading channels, for a bandwidth ratio of $k/n = 0.5$ with different number of layers. Notice that, although pilots are not explicitly transmitted, more feedback uses indirectly imply better channel estimation, improving the performance as the number of feedback transmissions increase.}
    \label{fig:fading}
\end{figure}

In order to explore the optimal number of layers in the case of a fading channel, we propose an experiment similar to the one in Figure~\ref{fig:layer_psnr}, in which the performance for a fixed bandwidth ratio ($k/n = 0.5$) is experimented with different number of layers. Results can be seen in Figure~\ref{fig:fading}.

Our results show firstly that the performance over the fading channel can be significantly improved compared to the results presented in Figure \ref{fig:snr_1fb_fading} with $L=2$ layers.
We also observe that we benefit from increasing the number of layers beyond $L=4$, as opposed to the AWGN channel. This can be explained by the fact that, in the case of a fading channel, feedback conveys information not only about the corruption/ noise over the channel, but it also contains information regarding the channel condition, and can be used for channel estimation. More accurate channel state estimation results in more accurate transmissions. 

\subsection{Noisy Feedback}\label{ss:noisy_feedback}

Finally we consider the impact of noisy channel output feedback, i.e., $\sigma_f^2 >0$. This is a particularly challenging setting as the encoder cannot track the quality of the receiver's reconstruction accurately, and hence, it cannot steer the decoder to the right decision as efficiently as possible. Indeed it is known that the known schemes with theoretical performance analysis, e.g., the Schalkwijk-Kailath scheme \cite{Schalkwijk:IT:67, Kailath:PIEEE:67}, break down even with a slightly noisy feedback.
While it is shown in \cite{Hyeji:NIPS:2018} that the proposed DNN-based feedback channel code provides some level of robustness against noise in the feedback channel, to the best of our knowledge, this is the first practical JSCC scheme that is robust to feedback channel noise.

\begin{figure}
    \centering
    \includegraphics[width=0.7\linewidth]{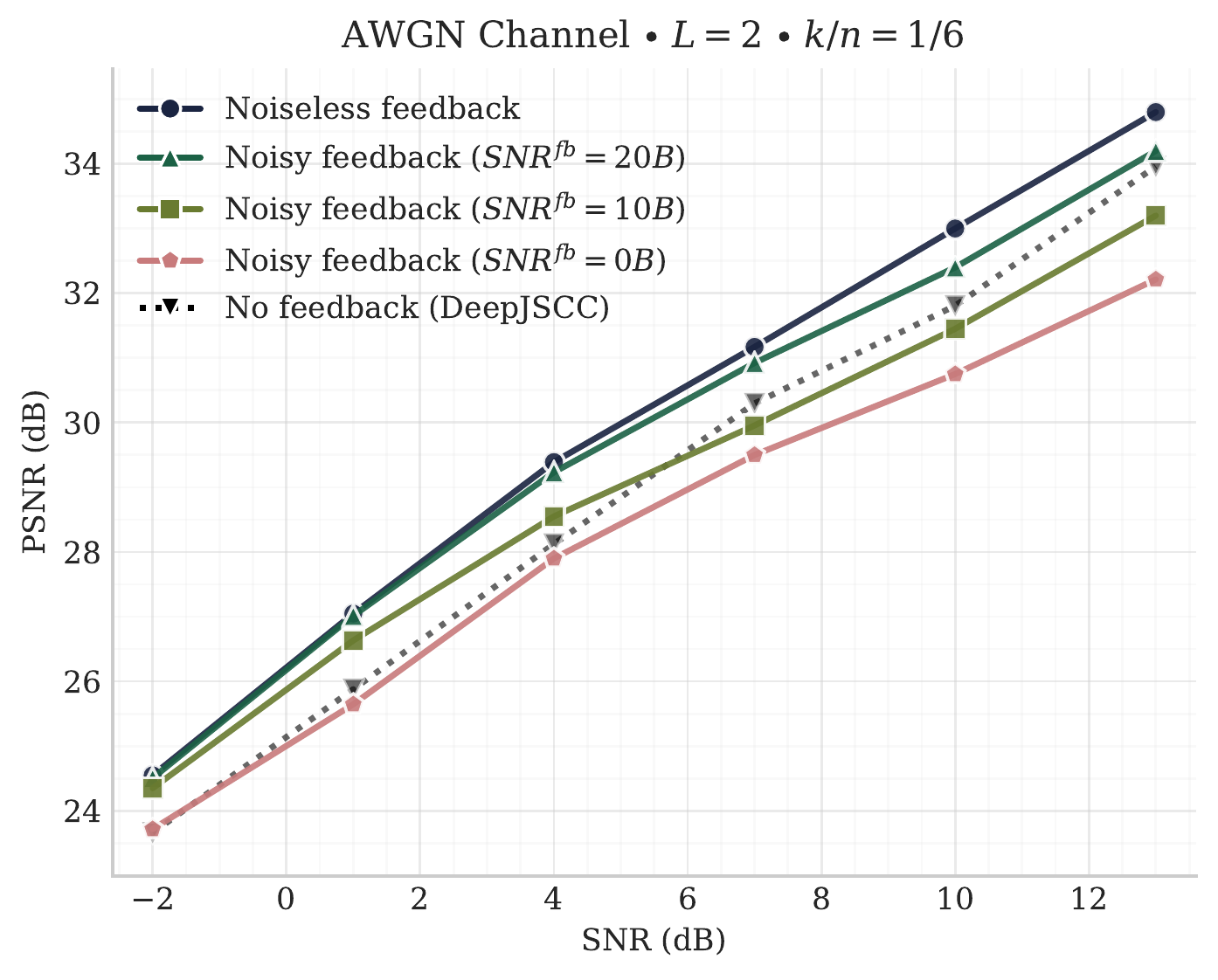}
    \caption{Model performance when feedback channel has noise. Models trained in low forward SNRs are more robust and can benefit from feedback even when $\SNR^{fb}$ is low.}
    \label{fig:noisy_snr}
\end{figure}

Figure~\ref{fig:noisy_snr} shows the performance for different feedback channel SNRs ($0$~dB, $10$~dB, $20$~dB, and noiseless) and the single layer model (no feedback). The SNR of the feedback channel is measured in terms of the channel input, i.e., $1/ \sigma_f^2$. We can see that our model is robust to noise in the feedback channel. When $\SNR^{fb}=20$~dB (high quality feedback), the performance is only slightly below the noiseless feedback transmission. As SNR decreases, the performance degrades, but overall remains quite high and competitive. The network can learn to make good use of the feedback even with $\SNR^{fb} = 10dB$ when forward channel SNR is low.

When the feedback channel is very noisy ($SNR^{fb} \! = \! 0$ dB), the transmission of additional layers still positively contributes to the refinement of the reconstruction, but the overall performance is below than what can be achieved by a single transmission (i.e., $L=1$). However, a hybrid approach could be proposed, in which depending on the noise level, the encoder decides either to use or not the channel output feedback information at each transmission, transmitting only refinement information instead of error correction. Such approach could enable that performance to be at least as good as the single transmission case, regardless of the feedback channel noise \cite{Kurka:ICASSP19}. An implementation of such hybrid solution is left for future work.

\subsubsection{Feedback Channel Mismatch}

Similarly to the mismatch between the training and test SNRs of the forward channel studied in Subsection \ref{ss:graceful_noiseless}, we want to study the effects of the variations in the feedback channel quality during deployment.

\begin{figure}
    \centering

    \subfloat[0dB]{%
        \includegraphics[width=0.49\textwidth]{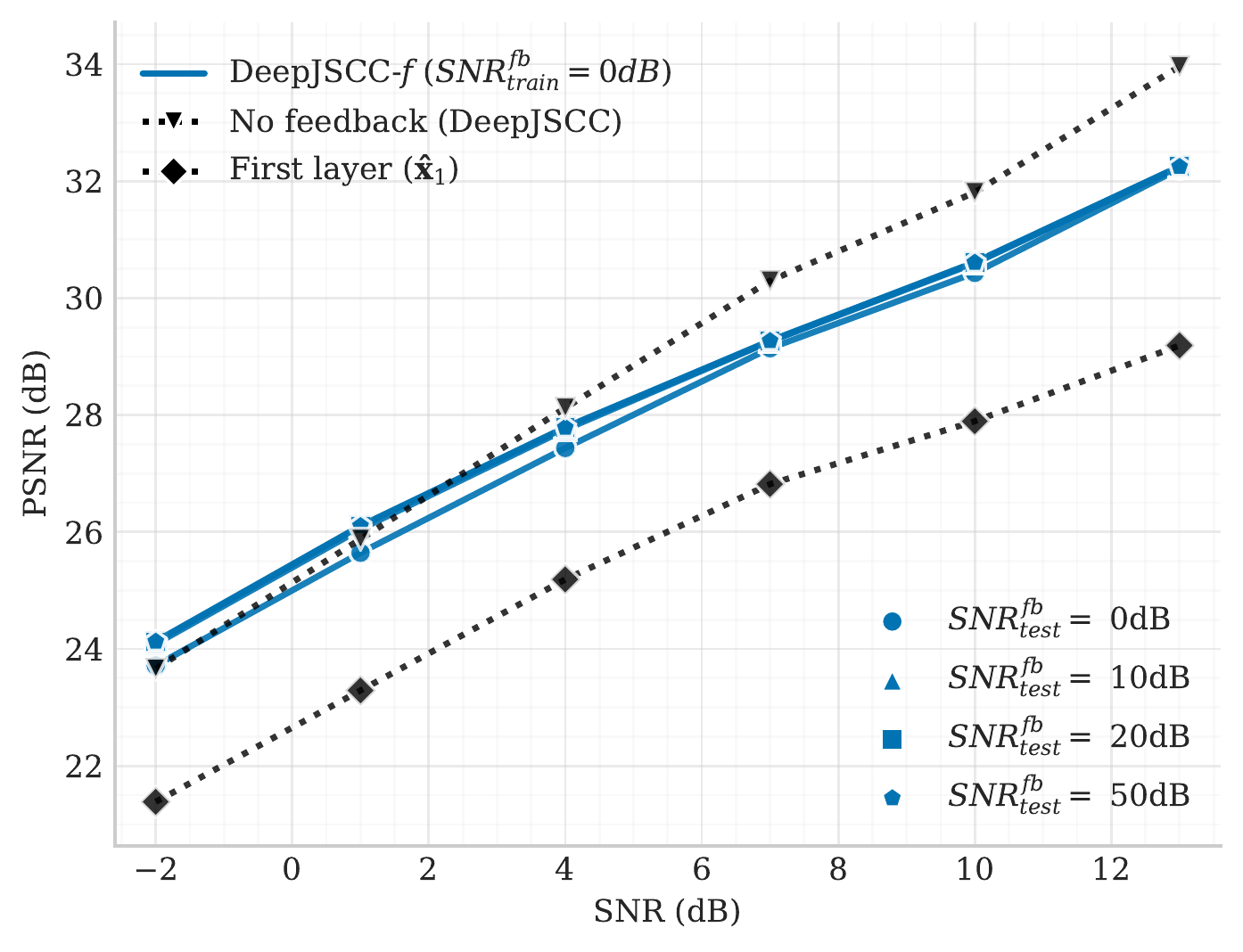}
        \label{fig:grfb0}}
    \hfill
    \subfloat[10dB]{%
        \includegraphics[width=0.49\textwidth]{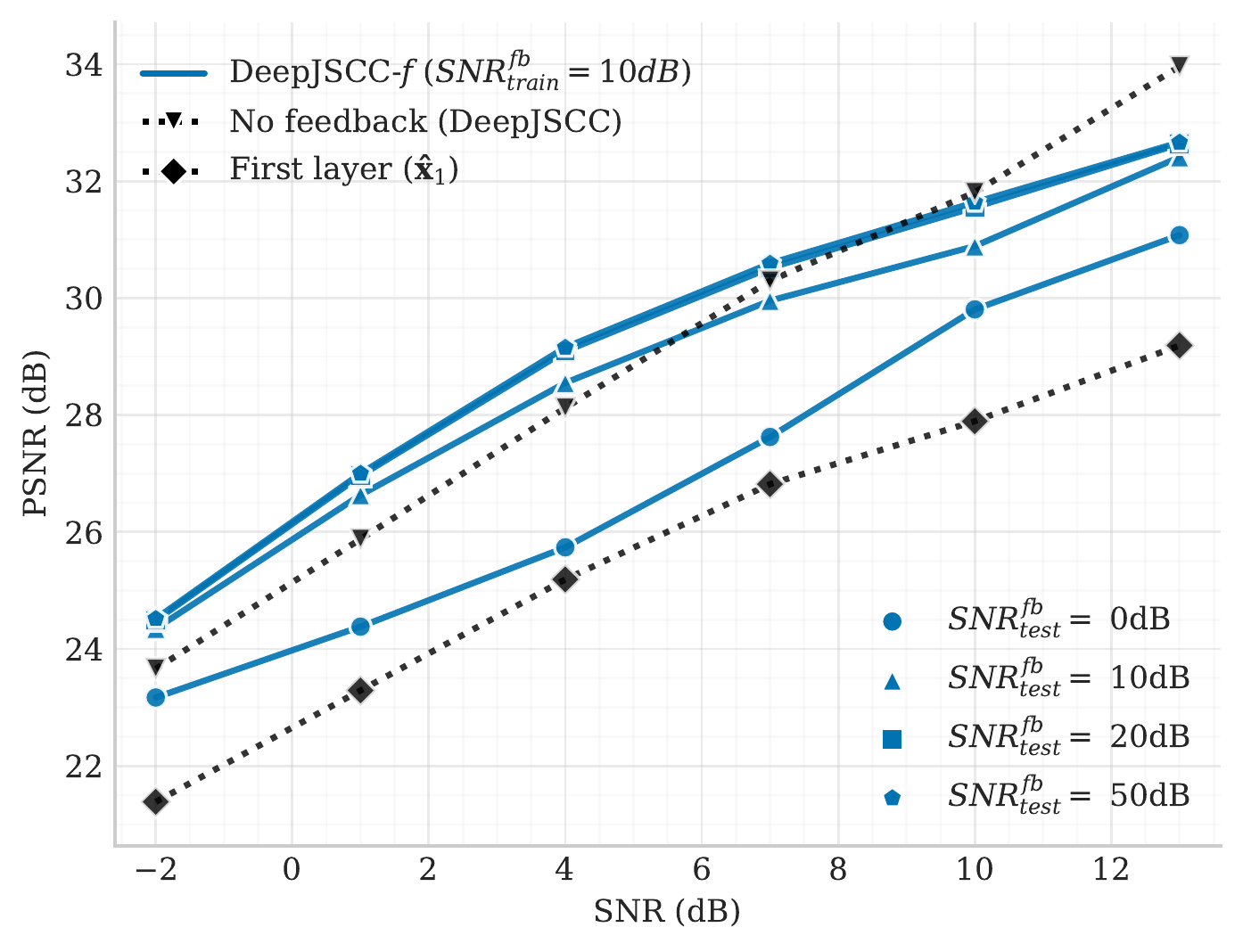}
        \label{fig:grfb10}}
    \\
    \subfloat[20dB]{%
        \includegraphics[width=0.49\textwidth]{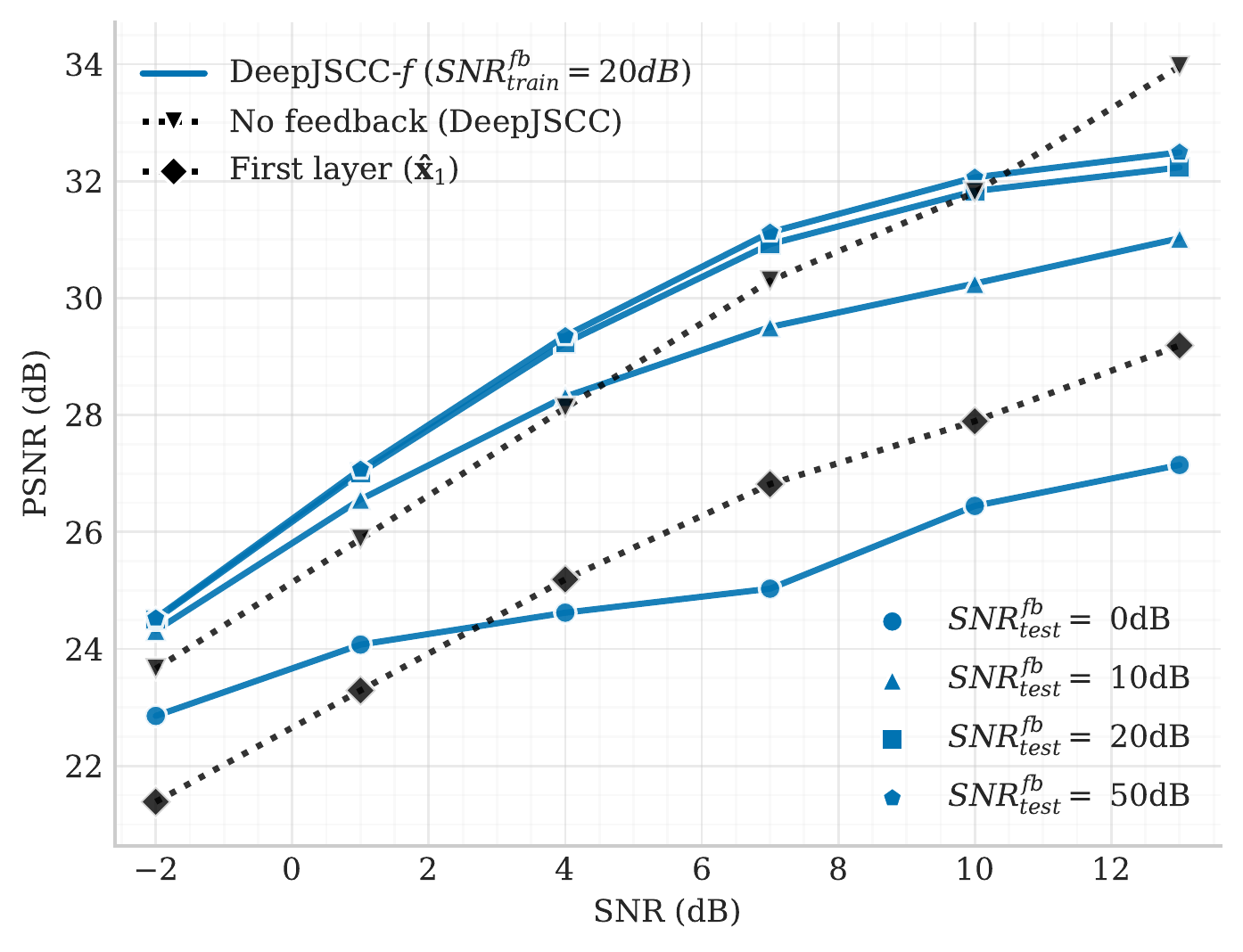}
        \label{fig:grfb20}}
    \subfloat[Noiseless]{%
        \includegraphics[width=0.49\textwidth]{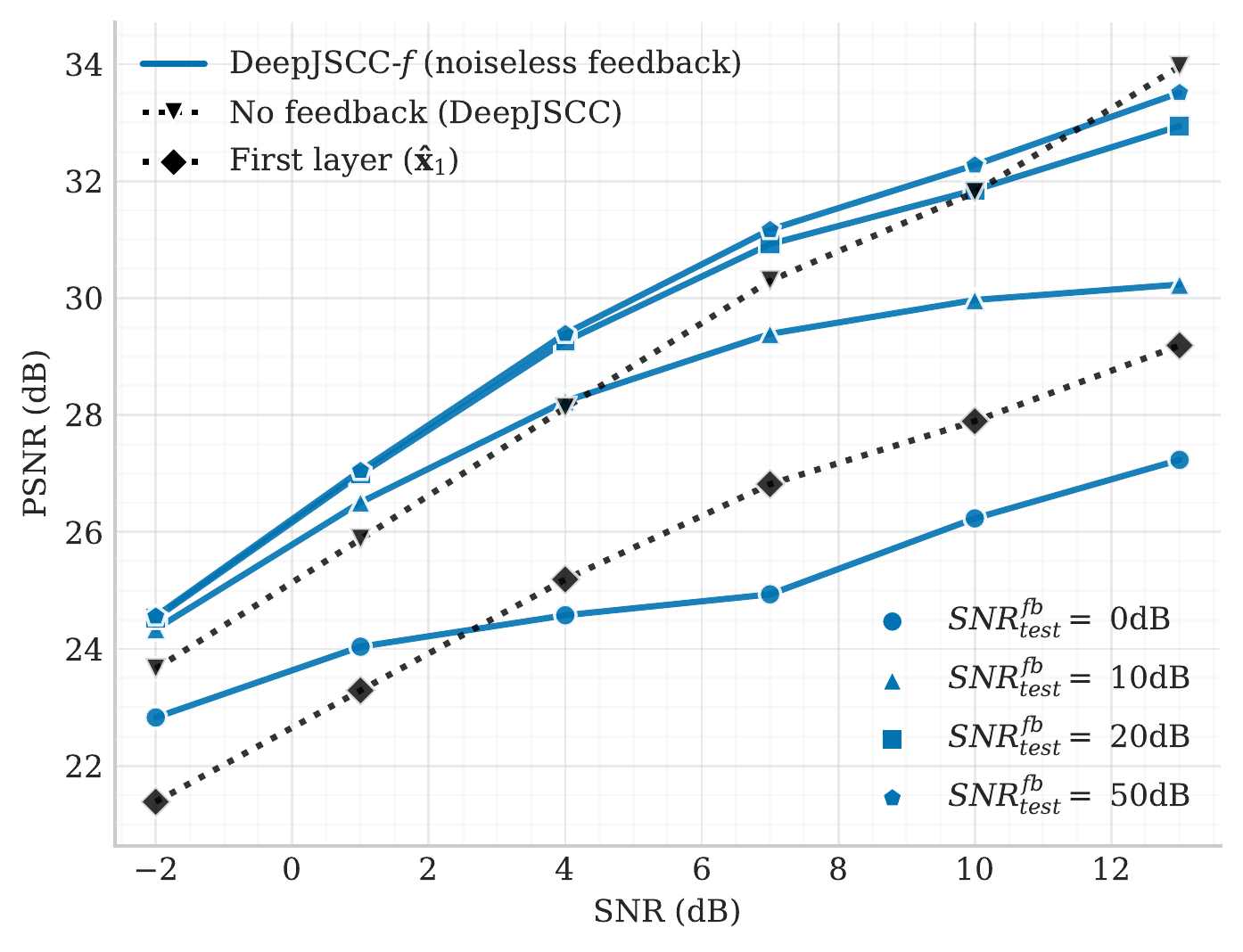}
        \label{fig:grfbinf}}
    \caption{Effect of the feedback channel degradation in models trained for different feedback channel SNRs. Although the performance deteriorates when the feedback channel quality decreases, it does progressively, presenting graceful degradation with respect to the feedback channel quality. Even in highly compromised channels, images can still be conveyed and communication is refined with successive transmissions.}\label{fig:grfb}
\end{figure}

Figure~\ref{fig:grfb} shows how models trained with feedback channel SNR equal to $0dB$, $10dB$ and $20dB$, and even noiseless feedback, perform when tested over different qualities of feedback channel, at different forward channel qualities, for the same models from Figure~\ref{fig:noisy_snr} ($L=2$, $k/n=1/6$).  We used AWGN for both forward and feedback channels. As baselines, we also plot the performance of a forward transmission scheme with DeepJSCC using the same bandwidth ratio as the complete transmission with feedback (i.e., $L=1$), and the partial performance of the base layer ($\bm{\hat{x}}_1$) with bandwidth ratio $k_1/n = 1/12$. The latter is included to see how much improvement the second layer brings despite the mismatch between the training and test feedback channel SNRs.

We see that, even in the cases on which the channel conditions differ drastically from the model designed, the feedback information can still be useful, improving the performance of the base layer. Remarkably, even the model trained without feedback noise can still perform reasonably well even at $\SNR^{fb} = 10dB$, especially in low forward channel SNRs, still surpassing the single transmission bound. Only in extremely adverse and disparate feedback channel conditions ($\SNR^{fb}_{train} - \SNR^{fb}_{test} \geq 20dB$) and high forward channel SNRs, sending the extra layer decrease the communication performance. However, even in those cases, the retrieved signal can still produce an identifiable picture (the PSNR of a failed transmission is approximately $15dB$). 

This shows that, as with the forward channel, \mbox{DeepJSCC-$f$} can present graceful degradation with the feedback channel quality as well, and can operate reasonably well in adverse scenarios that would completely compromise conventional digital systems.
On the other hand, the increase in the feedback channel quality has only a small impact in the performance, being slightly beneficial.

\subsection{Model Generalisation to Bigger Datasets}

The results presented in the previous sections considered models trained and evaluated with (distinct) images from the CIFAR10 dataset. Although CIFAR10 contains valid examples of colored images, they are low resolution images ($32\times32$ pixels) and depict only $10$ classes of objects (although our algorithm does not directly consider the class information in any way during its execution and all classes are transmitted equally). However, our scheme is not limited to a specific input size or type and, given its fully convolutional architecture, it can accept multiple dimension inputs.

Thus, to test our scheme on a more generic dataset, we trained it on the ImageNet dataset~\cite{imagenet_cvpr09}, a large dataset composed of 1.2 milliom images, with over 20000 classes. Training was performed in patches of size $128\times128$ and fed into the network in batches of $16$ images. We then evaluated our trained model in full resolution images of the Kodak\footnote{\url{http://www.r0k.us/graphics/kodak/}} dataset, with dimensions $768 \times 512$ pixels.

\begin{figure}
    \centering

    \subfloat[Original image ($\bm{x}$)]{%
        \includegraphics[width=0.49\textwidth]{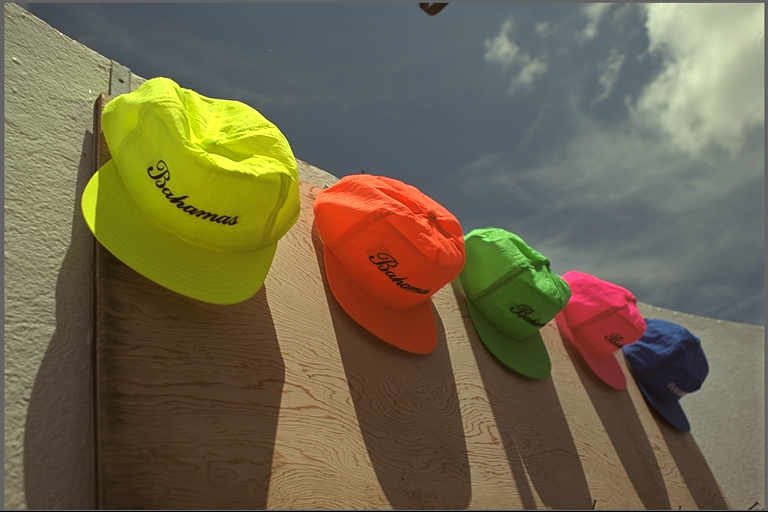}
        \label{fig:kdk_x_v3}}
    \hfill
    \subfloat[Base layer reconstruction ($\bm{\hat x}_1$)]{%
        \includegraphics[width=0.49\textwidth]{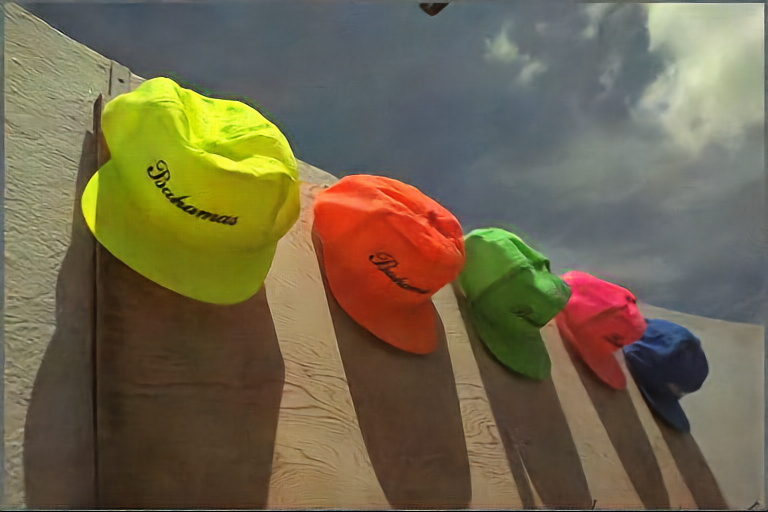}
        \label{fig:kdk_x1_v3}}
    \\
    \subfloat[Decoded image at second layer ($\bm{\hat{u}}_2$)]{%
        \includegraphics[width=0.49\textwidth]{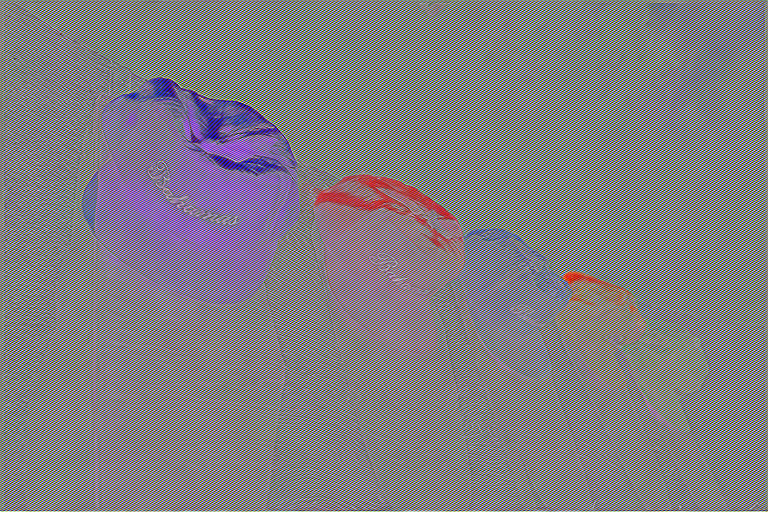}
        \label{fig:kdk_r2_v3}}
    \subfloat[Final reconstruction ($\bm{\hat x}_2$)]{%
        \includegraphics[width=0.49\textwidth]{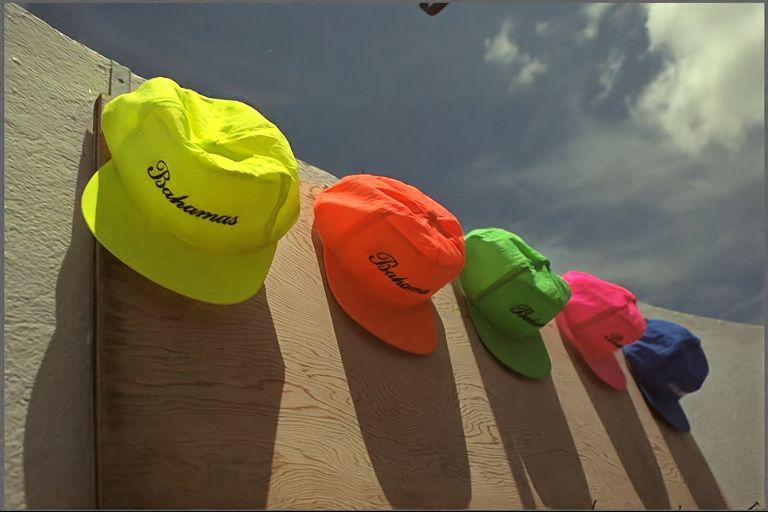}
        \label{fig:kdk_x2_v3}}
    \caption{Example of reconstructions obtained in different stages of the \mbox{DeepJSCC-$f$} scheme, for from model trained with Imagenet images and evaluated with Kodak dataset at the AWGN channel with SNR=1dB.}\label{fig:kdk_v3}
\end{figure}

Figure~\ref{fig:kdk_v3} shows the output of \mbox{DeepJSCC-$f$} at different stages, while transmitting an image at an AWGN forward channel with $\SNR=1$~dB, noiseless feedback channel, $L=2$ and $k/n = 1/6$. At the top left (\ref{fig:kdk_x_v3}) we see the original image being transmitted ($\bm x$) and top right is the reconstruction after the first transmission ($\bm{\hat{x}}_1$). At the bottom, we first present at the left, the image received by the second decoder before input to the combiner ($\bm{\hat{u}}_2$). Note that $\bm{\hat{u}}_2$ contains a residual between $\bm{\hat{x}}_1$ and $\bm{x}$, creating a refinement on the image quality. Also note that the autoencoder learns this representation in an unsupervised way, as the encoder receives only $\bm{x}$ and $\bm{\hat{x}}_1$, without any explicit subtraction operation. Finally, by combining the residual $\bm{\hat{u}}_2$ with $\bm{\hat{x}}_1$ through the combiner network, the final enhanced reconstruction $\bm{\hat{x}}_2$ is achieved as the output.

We also present, in Appendix~\ref{sec:images}, a visual comparison between \mbox{DeepJSCC-$f$} and separation-based schemes for sample images of the Kodak dataset. Our results (Figures~\ref{fig:visual_river} and \ref{fig:visual_house}) show that \mbox{DeepJSCC-$f$} can produce higher quality reconstructions, with increased details and better performance in all metrics considered (PSNR, SSIM, MS-SSIM).

\section{Conclusions}

In summary, this work presented, to the best of the authors' knowledge, the first practical implementation of a JSCC scheme exploiting channel output feedback -- \mbox{DeepJSCC-$f$} -- for images. By exploring the channel output feedback, we have shown that the scheme can achieve considerable gains in performance, when compared to a series of other schemes, including (a) JSCC without feedback; (b) state-of-the-art image compression codecs followed by practical and high performing channel codes; and (c) ideal capacity achieving channel codes.

Our experiments reveal that the use of the feedback channel improves the quality of the transmission, justifying the adoption of a multi-step strategy for image transmission, in which a source is sent over multiple layers, exploiting the feedback between transmissions. Moreover, flexible strategies such as variable rate transmission is also enabled, allowing a considerable economy of resources when transmitting in order to achieve a target quality goal.

Apart from the direct benefits of exploiting the feedback information, we also show that \mbox{DeepJSCC-$f$} has many other advantages, such as presenting analog behaviour and graceful degradation in case of variations in the channel quality, being able to adapt to either forward or feedback channel variations. Also, it can adapt to other channels, such as fading channel and to different datasets with diverse input file sizes.

This work is just a first step towards JSCC strategies using deep neural networks. Advances in the model architecture and training strategy can bring further performance improvements in the image transmission task, and the general design principles presented in this work can be potentially applied to other information sources, such as audio or video.

\bibliographystyle{IEEEtran}
\bibliography{references}

\newpage

\appendices

\section{Separation-based Scheme Used for Comparison}\label{sec:ldpc}

To compare the performance of our proposed algorithm to practical separation-based digital schemes, we consider a setup that uses different well established image compression codecs followed by low density parity check (LDPC) codes for error correction. For image compression, we consider JPEG, JPEG2000 and BPG, and we discount the header information for BPG and JPEG2000 when computing bit rates and transmission sizes for fair comparison. For the channel code, we consider all possible combinations of (4096, 8192), (4096, 6144) and (2048, 6144) LDPC codes (which correspond to 1/2, 2/3 and 1/3 rate codes) with BPSK, 4-QAM, 16-QAM and 64-QAM modulation schemes.

For each channel code configuration, we can define the maximum rate $R_{max}$ (bits per pixel) at which we can transmit an image (using the channel code rate), and empirically evaluate the frame error rate $\epsilon$ for each channel model and condition we consider. Then, we compress the images (using the different codecs) at the largest rate $R$ that satisfies $R \leq R_{max}$. We consider that the transmission can either be successful or fail, with probability of failure $\epsilon$.  When the transmission fails, we consider that the reconstruction at the receiver is set to the mean value for all the pixels per colour channel. When the transmission is successful the distortion is dictated by $R$ and the codec used. We then measure the average performance over the evaluation dataset.

\begin{figure}
    \centering
    \includegraphics[width=0.9\linewidth]{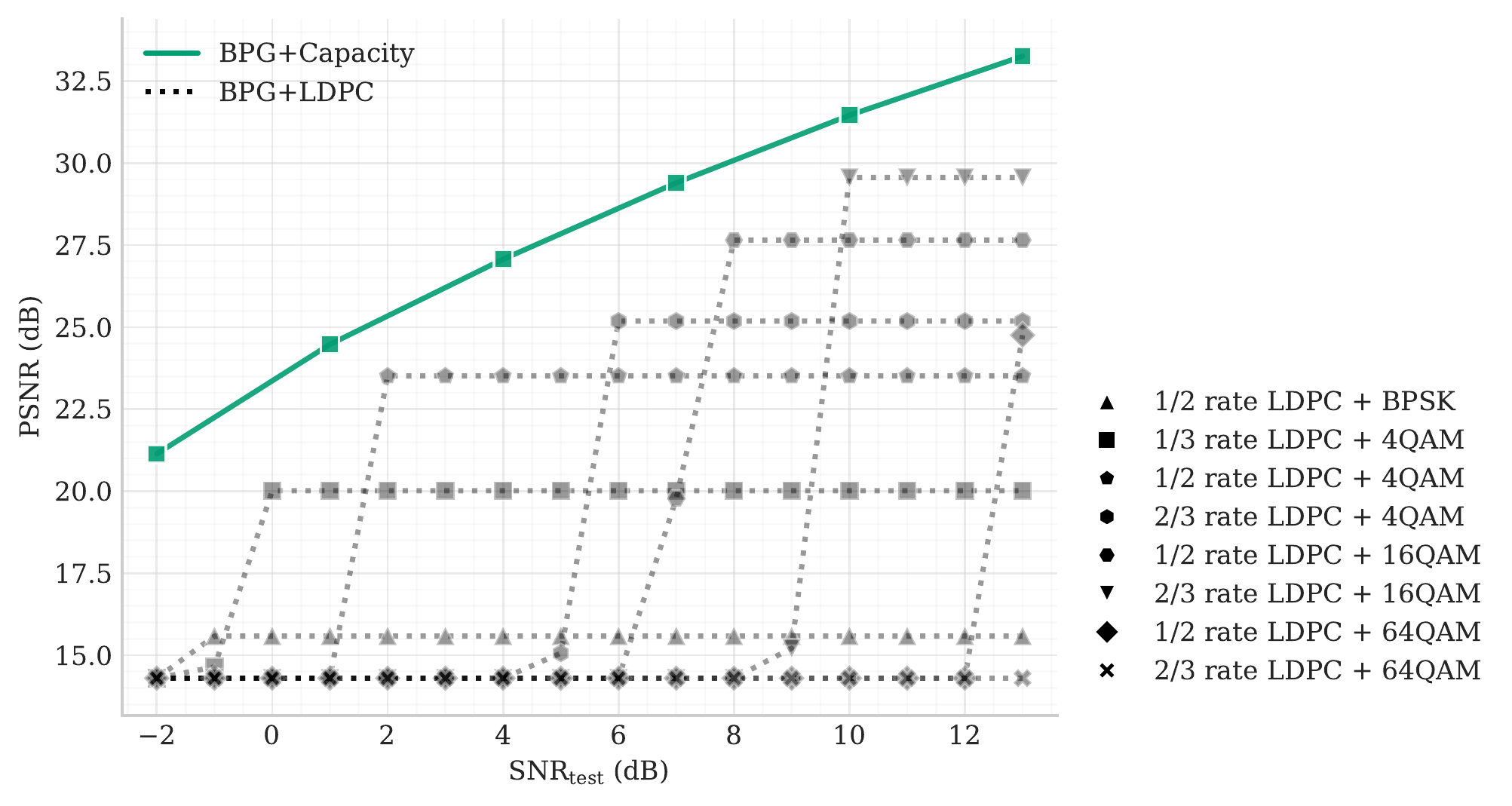}
    \caption{LDPC codes considered in this work. Capacity achieving bound also plotted for reference.}
    \label{fig:ldpc}
\end{figure}

Figure \ref{fig:ldpc} shows the individual results obtained from different configurations of code rates and modulations, for the $k/n = 1/6$ case. On the plots presented throughout the paper, we simply plot the envelope of the best performing values across all configurations.

\section{Example of Image Reconstructions from Different Schemes}\label{sec:images}

\begin{figure*}
  \begin{center}
\begin{tabular}{cc}

\textbf{Original Image} & \\
\includegraphics[width=0.49\textwidth]{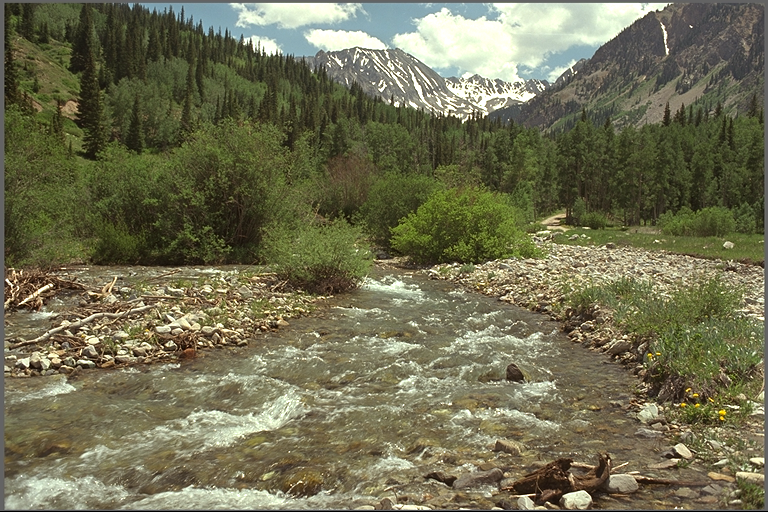} & \\
\textbf{JPEG+LDPC} & \textbf{JPEG2000+LDPC} \\
\includegraphics[width=0.49\textwidth]{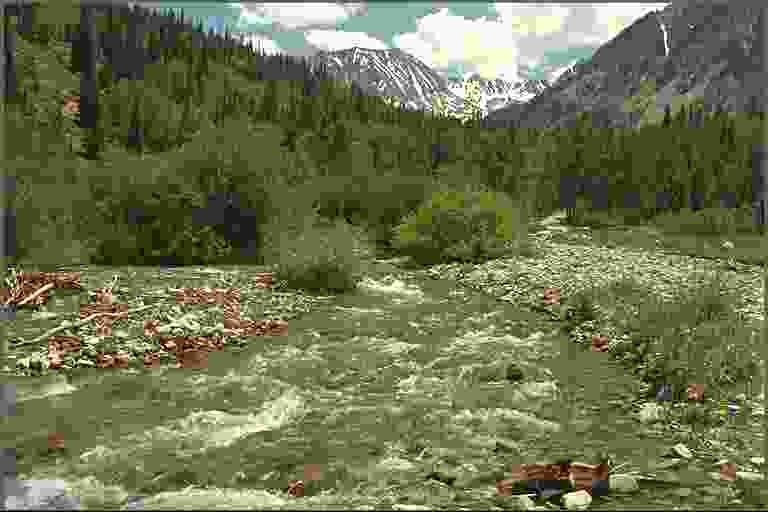} &
\includegraphics[width=0.49\textwidth]{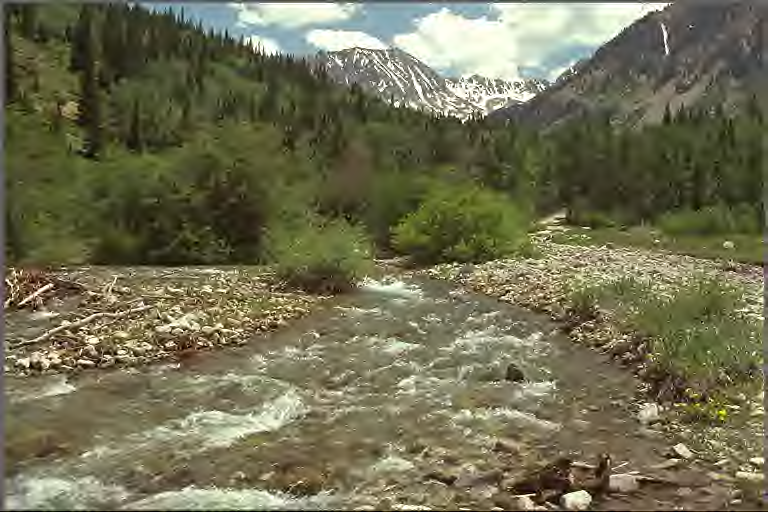} \\
PSNR: 18.97dB; SSIM: 0.383; MS-SSIM: 0.668 &
PSNR: 22.02dB; SSIM: 0.508; MS-SSIM: 0.830 \\
\\

\textbf{BPG+LDPC} & \textbf{\mbox{DeepJSCC-$f$}} \\
\includegraphics[width=0.49\textwidth]{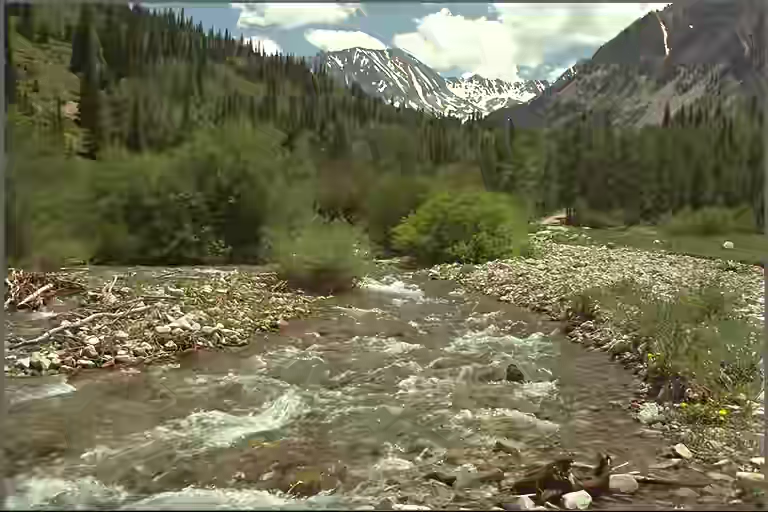} &
\includegraphics[width=0.49\textwidth]{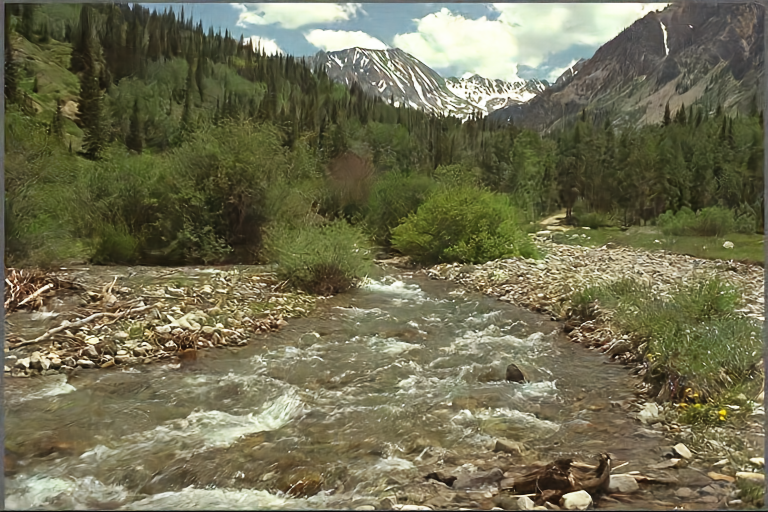} \\
PSNR: 22.98dB; SSIM: 0.539; MS-SSIM: 0.851 &
PSNR: 25.99dB; SSIM: 0.781; MS-SSIM: 0.948 \\

\end{tabular}
 \end{center}
    \caption{Comparison of reconstructed images from different schemes, including \mbox{DeepJSCC-$f$}. AWGN channel, compression rate $k/n=1/6$. Note how high frequency components (bushes' leaves, water waves) better preserve their details in \mbox{DeepJSCC-$f$}.}
    \label{fig:visual_river} 
\end{figure*}

\begin{figure*}
  \begin{center}
\begin{tabular}{cc}

\textbf{Original Image} & \\
\includegraphics[width=0.49\textwidth]{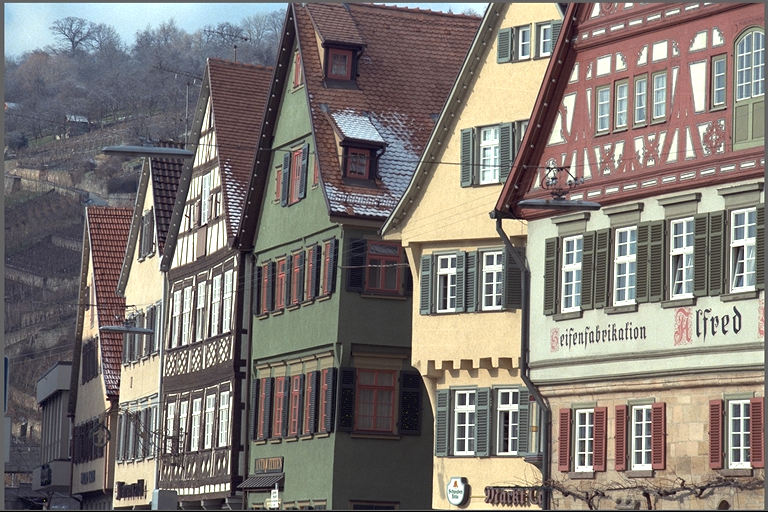} & \\
\textbf{JPEG+LDPC} & \textbf{JPEG2000+LDPC} \\
\includegraphics[width=0.49\textwidth]{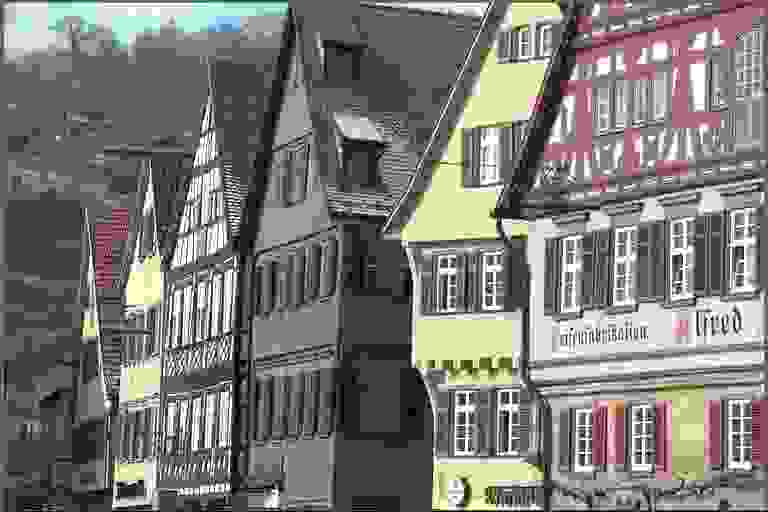} &
\includegraphics[width=0.49\textwidth]{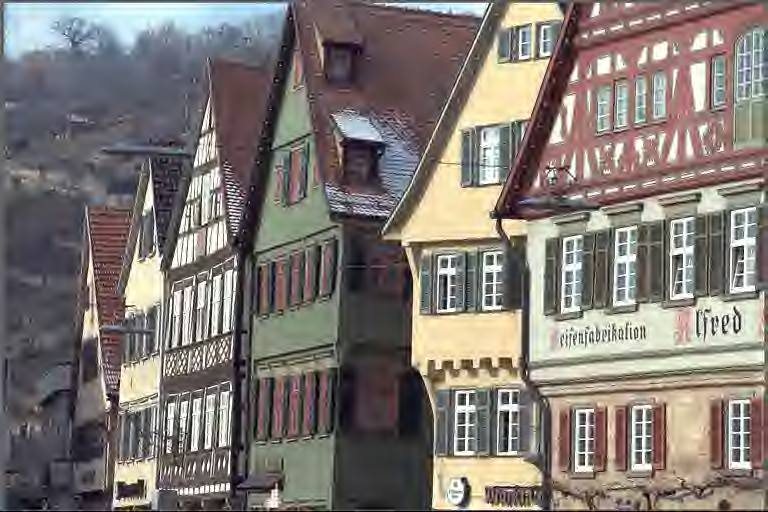} \\
PSNR: 18.97dB; SSIM: 0.383; MS-SSIM: 0.668 &
PSNR: 22.02dB; SSIM: 0.508; MS-SSIM: 0.830 \\
\\

\textbf{BPG+LDPC} & \textbf{\mbox{DeepJSCC-$f$}} \\
\includegraphics[width=0.49\textwidth]{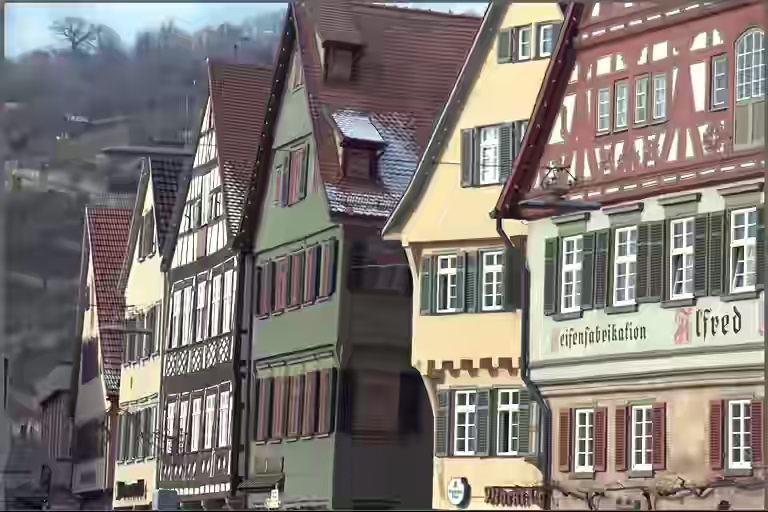} &
\includegraphics[width=0.49\textwidth]{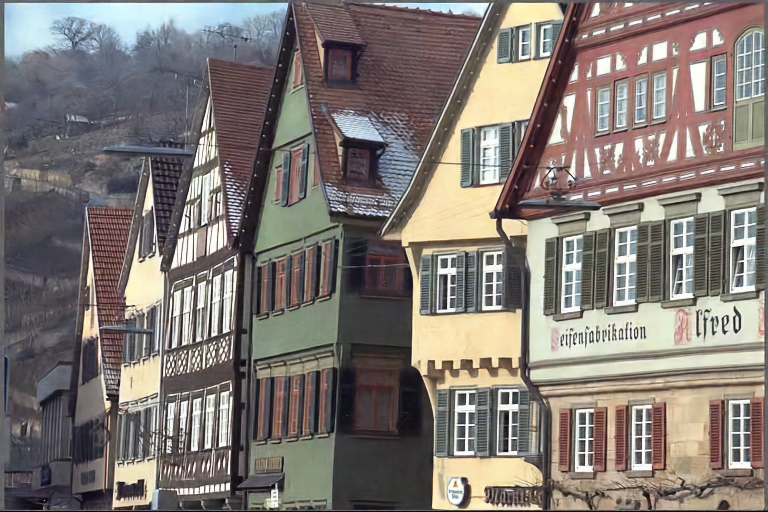} \\
PSNR: 22.98dB; SSIM: 0.539; MS-SSIM: 0.851 &
PSNR: 25.99dB; SSIM: 0.781; MS-SSIM: 0.948 \\

\end{tabular}
 \end{center}
    \caption{Comparison of reconstructed images from different schemes, including \mbox{DeepJSCC-$f$}. AWGN channel, compression rate $k/n=1/6$.\mbox{DeepJSCC-$f$} presents superior performance in all metrics considered.}
    \label{fig:visual_house} 
\end{figure*}

In order to provide a visual comparison of our algorithm versus traditional digital separation-based schemes, we present sample images from the Kodak dataset, produced on a setup on AWGN channel, SNR=1dB and bandwidth compression $k/n = 1/6$. We compare \mbox{DeepJSCC-$f$} (noiseless feedback) with separation-based scheme using 1/3 rate LDPC + 4QAM as channel code and JPEG, JPEG2000 and BPG as source codes. Results can be seen in Figures~\ref{fig:visual_river} and \ref{fig:visual_house}, where the PSNR, SSIM and MS-SSIM performances are computed.

Note that, for the images considered, \mbox{DeepJSCC-$f$} present superior performance in all metrics when compared to the sepration-based schemes. This gain is particularly noticed in high frequency elements (bushes, leaves and water in Figure~\ref{fig:visual_river}; background landscape, house details in Figure~\ref{fig:visual_house}).

Also note these results correspond to \mbox{DeepJSCC-$f$}  optimised for MSE as loss function. It is possible to optimise \mbox{DeepJSCC-$f$} directly to other metrics (e.g. SSIM or MS-SSIM), producing even superior results than the presented for the referred metrics.

\end{document}